# The Interaction of Supernova Remnant G357.7+0.3 with the Interstellar Medium

J.P. Phillips, R.A. Marquez-Lugo

Instituto de Astronomía y Meteorología, Av. Vallarta No. 2602, Col. Arcos Vallarta, C.P. 44130 Guadalajara, Jalisco, México   e-mail : jpp@astro.iam.udg.mx

# 1. Abstract

The supernova remnant (SNR) G357.7+0.3 appears to have caused considerable shredding of the local interstellar medium (ISM), leading to the formation of multiple cloud fragments having bright rims and cometary structures. We investigate five of these regions using mid-infrared (MIR) imaging and photometry deriving from the Spitzer Space Telescope (*SST*), as well as photometry deriving from the 2MASS near-infrared all sky survey, the Mid-Course Science Experiment (MSX), and the Multiband Imaging Photometer for Spitzer (MIPSGAL) survey of the Galactic plane. It is noted that two of the rims show evidence for emission by shock excited $H_2$ transitions, whilst the centres of the clouds also show evidence for dark extinction cores, observed in silhouette against the bright emission rims. Levels of extinction for these cores are determined to be of order $A_V \sim$ 17-26 mag, whilst densities n(HI) are of order $\sim 10^4$ cm$^{-3}$, and masses in the region of ~40-100 $M_\odot$. It is shown that the wavelength dependence of extinction is probably similar to that of Cardelli et al. and Martin & Whittet, but differs from the MIR extinction trends of Indebetouw et al. The distributions of Class I young stellar objects (YSOs) implies that many of them are physically associated with the clouds, and were likely formed as a result of interaction between the clouds and SN winds. A determination of the spectral energy distributions (SEDs) of these stars, together with 2-D radiative transfer modelling of their continua is used to place constraints upon their properties.

**Key Words:** stars: formation --- (stars:) supernovae: individual: G357.7+0.3 --- ISM: kinematics and dynamics --- ISM: jets and outflows --- ISM: clouds --- (ISM:) supernova remnants

# 1. Introduction

The expansion of supernovae remnants (SNRs) into the interstellar medium (ISM) is believed to give rise to a broad range of consequences, both for the SNRs themselves and IS environments in which they are immersed. Velocities of shell expansion tend to be reduced as a result of the sweeping-up of interstellar gas (see e.g. Van der Swaluw 1971), for instance, whilst the envelopes may be distorted through interaction with denser regions of the ISM, and prone to "blow-outs" (i.e. increased rates of expansion) where the ISM is hotter and less dense (see e.g. Caswell 1977, Caswell et al. 1983; and Gray 1994 for specific examples of these phenomena). Globules located close to the SNRs are likely to be shredded and destroyed where velocities are greater than ~50 km s$^{-1}$ (e.g. Boss 1995; Foster & Boss 1996, 1997; Vanhala & Cameron 1998; Fukuda & Hanawa 2000), a situation which is likely to occur within ~10 to 100 pc of the supernova event, depending upon the structure of the ISM, and the evolutionary status of the pre-impact cores (e.g. Oey & García-Segura 2004; Vanhala & Cameron 1998). More distant inhomogeneities will tend to be subject to shock compression and heating, leading to gravitational instability and the formation of new stars (e.g. Boss 1995; Foster & Boss 1996, 1997; Nakamura et al. 2006; Melioli et al. 2006; Boss et al. 2008).

Evidence for such interactions is both common and ubiquitous, and includes the observation of shock excited molecular and ionic transitions (see e.g. Burton et al. 1988; Richter, Graham & Wright 1995; Rosado et al. 2007); enhanced emission from the 1720 MHz ($^2\Pi_{3/2}$, J = 3/2, F = 2→1) maser line of OH (e.g. Wardle & Yusef-Zadeh 2002, and references therein); broad CO, OH and other molecular transitions (see e.g. Arikawa et al. 1999, Reach et al. 2005 (W28); Reach & Rho 1999, Reach et al. 2002 (3C 391); Seta et al. 1998, and Reach, Rho & Jarrett 2005 (W44)); the formation of young stellar objects (YSOs) about the edges of the SNRs, and evidence for bright frontal structures and modified cloud structures.

An example of such an SNR/ISM interaction has recently been noted for the case of G357.7+0.3, where Phillips et al. (2009) found that many of the clouds have a ragged and/or cometary appearance, and possess linear structures, tails, and bright emission rims pointing directly towards the SNR shell. Evidence for interaction is also provided through the detection of OH 1720 MHz emission close to the rim of the SNR (Yusef-Zadeh et al. 1999); a likely indicator of interaction with the ISM (see the references cited above). Only one of these clouds (G357.46+0.60) is seen to be directly interacting with the SNR itself, and shows evidence for a bright emission rim, ram-pressure stripping of globule material, and a Class I YSO close to the HI/HII interface. Most of the other fragments appear to be located outside of the SNR, although their morphologies are again consistent with interaction with a SN wind (see our more detailed description in Sect. 3). It therefore seems possible that the ISM is interacting with pre-cursor winds, generated prior to the primary SN event, or that the SNR shell is much larger than would be supposed from radio mapping of the shell.

One of the structures takes the form of a plume extending ~14 pc behind the Mira variable V1139 Sco. This may represent a case where the Mira is interacting directly with the SN winds, or arise due to the chance superposition of two unrelated phenomena. Phillips et al. (2009) noted that neither of these possibilities can entirely be discounted. It was also noted that the presence of Class I YSOs close to the bright emission rims suggested that SNR G357.7+0.3 was triggering star formation throughout this entire region.

We shall consider the detailed morphologies of several of these structures, as well as constraining the physical properties of the related YSOs. We shall note that two of the frontal structures appear to represent regions of shock excited $H_2$ emission, whilst several of the rims appear to be associated with enhanced regions of extinction; features which may represent the original cores of the globules silhouetted against bright shocked material, or regions where compression of neutral material is leading to higher column densities of dust. Estimates are made of the extinctions, densities and masses of these regions. Finally, we shall use the 2-D radiative transfer modelling of Robitaille et al. (2006) and the fitting tool of Robitaille et al. (2007) to analyse the NIR-MIR spectral energy distributions (SEDs) of YSOs located near regions of interaction. This will be used to place constraints upon the physical parameters of the stars, including their luminosities, masses, temperatures and radii.

## 2. Observations

We shall present observations of five compact IS clouds located within the region of interaction about G357.7+0.3, including photometry and imaging deriving from the near infrared (NIR) 2MASS all sky survey; the far infrared (FIR) Infrared Astronomical Satellite (IRAS); the mid-infrared (MIR) Midcourse Science Experiment (MSX); the MIR Galactic Legacy Infrared Midplane Survey Extraordinaire (GLIMPSE II), and an MIR survey of the Galaxy undertaken with the Multiband Imaging Photometer for Spitzer (MIPSGAL). Both of the latter surveys were undertaken using the Spitzer Space Telescope (*SST*). Details of the data bases employed, and procedures used in the analysis of the data can be found in Skrutskie et al. (2006) (2MASS); at the web site http://irsa.ipac.caltech.edu/IRASdocs/toc.html (IRAS); in Benjamin et al. (2003) (GLIMPSE); in Carey et al. (2009) (MIPSGAL); and in Mill et al. (1994) and Price et al. (1998) (MSX). We shall describe here only the most salient features of importance to this analysis.

The catalogues and mosaics of the GLIMPSE II project were published over a period extending between 2007 and the Spring of 2008, and cover approximately 25 square degrees of the Galactic plane within the regimes 350° < l < 10°, and -2° < b < +2°. Mapping was undertaken using the Infrared Array Camera (IRAC; Fazio et al. 2004), and employed filters having isophotal wavelengths (and bandwidths $\Delta\lambda$) of 3.550 $\mu$m ($\Delta\lambda$ = 0.75 $\mu$m), 4.493 $\mu$m ($\Delta\lambda$ = 1.9015 $\mu$m), 5.731 $\mu$m ($\Delta\lambda$ = 1.425 $\mu$m) and 7.872 $\mu$m ($\Delta\lambda$ = 2.905 $\mu$m). The spatial resolution varied between ~1.7 and ~2 arcsec (Fazio et al. 2004), and is reasonably similar in all of the bands, although there is a stronger diffraction halo at 8 $\mu$m than in the other IRAC bands. This leads to differences between the point source functions (PSFs) at ~0.1 peak flux. The maps were published at a resolution of 0.6 arcsec/pixel.

We have used the Spitzer GLIMPSE results to produce imaging of the clouds in the 5.8 and 8.0 $\mu$m IRAC bands. Fluxes were very much weaker at 3.6 and 4.5 $\mu$m, where results are also strongly affected by the field star contribution. We have therefore not included imaging at the shorter MIR wavelengths. We have also used the GLIMPSE II point source catalog (PSC), together with the diagnostic diagram of Allen et al. (2004) to identify Class I YSOs within the region of interaction. These are identified in the images using green and blue circles, depending upon whether they are included in the modelling to be described in Sect. 6.

The GLIMPSE II PSC contains about 19 million sources, and details of the selection criteria and processing are available at http://irsa.ipac.caltech.edu/data/SPITZER/GLIMPSE/doc/glimpse2_dataprod_v2.0.pdf. The sources are required to have two detections within any particular IRAC band, together with at least one detection in another photometric channel (which may include the other IRAC bands, or the 2MASS $K_S$ channel). This results in a 99.5 % level of source reliability (i.e. < 0.5 % of sources are spurious). The catalog is reasonably complete down to quite deep levels, of order magnitudes 14, 12, 10.5 and 9 at 3.6, 4.5, 5.8 and 8.0 $\mu$m, below which source count numbers decline precipitously, and photometric errors increase. Finally, the typical photometric accuracy is quoted as being < 0.2 mag, and we find the mean errors for our YSOs to vary from ~0.08 mag at 3.6 $\mu$m, through to 0.15 mag at 5.8 $\mu$m.

Whilst there is little doubt that the catalogue as a whole has a high level of reliability, it is always possible that contamination may arise as a result of local extended emission. An example of this is mentioned by Marquez-Lugo & Phillips (2010), where they find that frontal structures in G355.9-02.5 affect the photometry of YSOs. This is most likely, in our present case, to affect the probable triggered YSOs, the SEDs of which will be subject to more detailed modelling. Our analysis suggests that this is unlikely to affect our overall conclusions, however, or distort photometry to an unacceptably large degree (say > 30%).

Finally, we have investigated the MIPSGAL 24 $\mu$m fluxes of those YSOs we believe to have been triggered by the SNR, a process which involves identifying the fluxes of the stars with 14 arcsec circular apertures. Since all of these stars are located in regions of higher surface brightness emission, it has also been necessary to determine levels of background at six hexagonally located regions about the stars. These backgrounds are then subtracted from the results to yield an uncontaminated stellar flux.

Surface brightness profiles have been obtained in all four of the IRAC bands, and are composed of "background" components of flux, and the spatially more restricted emission deriving from the clouds themselves. The background arises from extended gas within the region of interaction, and unrelated emission along the line of sight, both of which are strong within the longer wave photometric channels. We have therefore removed a constant value of surface brightness from each of the profiles in order to emphasise changes associated with the clouds themselves. The observed surface brightness S(3.6) at 3.6 $\mu$m is represented in terms of S(3.6) = $S_{REV}$(3.6) + $\Delta$3.6, for instance, where $S_{REV}$(3.6) is the relative surface brightness illustrated in our profiles, and $\Delta$3.6 is the background component of emission. The values ($\Delta$3.6, $\Delta$4.5, $\Delta$5.8, $\Delta$8.0) are indicated in the captions to the figures. We have also used values of $S_{REV}$ to determine the flux ratios F(3.6$\mu$m)/F(5.8$\mu$m) and F(4.5$\mu$m)/F(8.0$\mu$m), although this is only possible where the 3.6 and 4.5 $\mu$m fluxes permit viable estimates of these parameters. These have been used to assess likely emission mechanisms within the clouds using the diagnostic colour diagram of Reach et al. (2006). Note that whilst ratios based upon surface brightnesses S are consistent with PAH band emission, they are not necessarily indicative of ratios within the clouds themselves.

Such a flux ratio analysis is necessarily somewhat approximate, given that estimates of $\Delta$3.6, $\Delta$4.5 & etc may be slightly in error; that shorter wave fluxes may be contaminated by stars; and that surface brightnesses are affected by scattering in the IRAC camera. This latter problem has been described in several previous papers (see e.g. Phillips et al. 2010), and may require corrections of as large as 0.944 at 3.6 $\mu$m, 0.937 at 4.5 $\mu$m, 0.772 at 5.8 $\mu$m and 0.737 at 8.0 $\mu$m. All three of these uncertainties may affect the colours of the sources, and lead to corresponding uncertainties in the MIR emission mechanisms. This results in ambiguities concerning the role of PAH emission, as is more fully described in Sect. 4.

In addition to the GLIMPSE analysis described above, the inner 258 square degree of the Galactic plane has been surveyed through the MIPSGAL project. Mapping was obtained at 24 $\mu$m and 70 $\mu$m using the Multi-band Imaging Photometer for Spitzer (MIPS), the seminal paper for which is provided by Rieke et al. (2004). The 24 $\mu$m pixel sizes are 1.25 arcsec, and the spatial resolution of the results is 6 arcsec. Here again, we provide 24 µm mapping of the region of interaction about SNR G357.7+0.3 (see Figs. 1 & 2), and use the results to provide comparative imaging of each of the clouds

The 2MASS all-sky survey was undertaken between 1997 and 2001 using 1.3 m telescopes based at Mt Hopkins, Arizona, and at the CTIO in Chile. We have used these results to evaluate fluxes for several probable examples of triggered YSOs, and employed these in the modelling in Sect. 6.

IRAS was launched in January 1983, and tasked with mapping 96% of the sky in wavebands at 12 µm, 25 µm, 60 µm and 100 µm. The resolution ranged from 30 arcsec at 12 µm through to 2 arcmins at 100 µm. The PSC has been used to identify likely YSOs within the clouds investigated here, and investigate the possible association of these sources with Class I YSOs.

Finally, various of our sources were observed during the MSX mission, and this permitted observations at 8.28 µm, 12.13 µm, 14.65 µm and 21.3 µm. Photometry from this catalogue was used to define the spectra of several YSOs.

### 3. The Fragmentation Zone about SNR G357.7+0.3, and the Identification and Location of the YSOs

A 24 µm image of the ISM about G357.7+0.3 is illustrated in Fig. 1, based upon results deriving from the MIPSGAL project. Most of the region of wind-swept structures is confined to a rectangular area to the upper right-hand side, the approximate limits of which are defined by the two white lines. We have also included an image of the source at 834 MHz (Gay 1994, indicated by green), contour mapping at 2695 MHz (Reich & Fürst 1984; white contours), and contours of 1720 MHz maser emission deriving from Yusef-Zadeh et al. (1999) (yellow contours). It is apparent from this that the easternmost clouds (labelled A & B in Fig. 2, and including the cometary structure at $l = 357.46°$, $b = +0.60°$) appear to point directly towards the SNR. They also have bright emission rims on the side of the SNR, and are located within ~30 arcmin of the limits of the shell. Given the presence of OH 1720 MHz emission in the outer regions of the G357.7+0.3, Phillips et al. (2009) suggested that the wind swept features were likely to be caused by the SNR – although the fact that they are located well outside of the main emission shell may imply that the SNR envelope is very much larger than appears the case. Alternatively, it may be possible that the clouds are interacting with pre-cursor SN winds, or with winds from a cluster of higher mass stars previously associated with the SN progenitor.

It is also possible that other mechanisms may be responsible for this distinctively fragmented regime, however, such as the winds and radiation pressures associated with HII regions. However, it is well known that HII regions have strong and distinctive emission signatures, and are very clearly visible within the IRAC wavelength regime (see e.g. Phillips & Perez-Grana 2009; Phillips & Ramos-Larios 2008). We see little evidence for such components in the present IRAC-MIPSGAL mosaics, although high and variable extinctions may help in masking such regimes. It may also be possible that radiation pressure and winds from a bright and massive cluster are similarly responsible for fragmenting the ISM. Here again however, we see no obvious candidates capable of causing the scale of fragmentation observed here, and it is far from clear whether this presents a plausible explanation for this present emission regime.

It therefore seems likely that the SNR is in some way responsible for these structures. Whilst other and less obvious mechanisms may also play a role, it is probable that the region has been sculpted by a single large-scale event, and that this has largely destroyed the local ISM. We shall see however that the scale of destruction varies over this regime, and there is evidence that it is somewhat reduced at larger distances from the SNR (see Sect. 4.5 below).

The various fragmented clouds are labelled A though E (see Fig. 2), and will be considered in further detail Sect. 4 below. The green circles, by contrast, correspond to Class I YSOs, and these are identified through their [3.6]-[4.5] and [5.8]-[8.0] colours in the GLIMPSE PSC.

There are various ways of identifying YSOs, and the differing evolutionary stages through which these stars pass. It is often the case that YSOs are defined as Class 0-III sources for instance, with the emission from Class 0 sources being dominant in the FIR, and arising from an infalling envelope of neutral gas and dust. Class I sources have strong emission from the outer envelopes and accretion disks (the latter particularly evident in the MIR), whilst Class II and III source have weaker (optically thick and thin) accretion disk emission, and much less evidence for longer wave excesses (see e.g. the early papers by André (1994), Lada & Wilking (1984), Adams et al. (1987), and Lada (1991)). This represents a less than perfect way of determining the state of evolution of the stars, however; differing orientations of protostars can lead to differing estimates of Class, even though their phases of evolution are very closely similar. It has therefore been proposed that YSOs are better classified in terms of an evolutionary "Stage", defined to accord more closely with their actual states of evolution.

An example of such an analysis is provided by Robitaille et al. (2006), where the use of [3.6]-[5.6] and [8.0]-[24.0] indices permits one to discriminate between differing "Stages" of evolution. Stage I YSOs (roughly the equivalent of Class I sources) have very large indices [8.0]-[24.0], for instance, a consequence of strong longer wave emission from the YSO envelopes. These indices are significantly greater than is the case for Stage II and III YSOs, and the discrimination between differing Stages of evolution is relatively clear. Excellent studies of YSOs in which such indices have been employed can be found in Reach et al. (2004), Muzerolle et al. (2004), and Rho et al. (2006). However, point source photometry is still not available for the MIPSGAL survey, and the number of 24 $\mu$m sources within the field is of order ~6000. We have therefore undertaken a slightly differing, two phase program of YSO analysis, in which we initially filter the sources using the IRAC PSC, and further investigate their status by employing an IRAC-MIPSGAL colour analysis.

Class I YSOs tend to have larger [3.6]-[4.5] and [5.8]-[[8-0] indices, and differing Classes of YSO occupy differing regimes of the IRAC colour plane (Allen et al. 2004). Whilst there may be some overlap between the Class I and Class II sources, the identifications are likely to have a reasonable level of verisimilitude. We have used this procedure to make an initial identification of Class I sources in the region of Fig. 2, as shown in the IRAC colour diagram illustrated in Fig. 3. We also show the reddening vector for $A_V$ = 20 mag (based on the extinction trends of Indebetouw et al. (2005)), and mean errors in the YSO colour estimates. It is clear from this that reddening and errors in the measured colours of the sources can cause some intermixing of the YSO Classes, and that a proportion of Class II sources may be erroneously classified as Class I.

It will finally be noted that we have identified sources having indices [5.8]-[8.0] > 1, [3.6]-[4.5] < 0.3 as Class I YSOs, although they extend somewhat below the modelling of Allen et al. (2004); it is likely that they represent Class I sources having lower envelope densities. Similarly, a small proportion of sources have larger indices [5.8]-[8.0] and [3.6]-[4.5], and lie above the regime of the Class I models. These are also likely to correspond to Class I sources, and are so interpreted here. Their location outside of the Class I regime may be due to errors in the photometry, or a consequence of the limited ranges of parameter considered by Allen et al. The inclusion of these sources makes little difference to our conclusions or results.

It is clear, from Fig. 3, that whilst the YSOs are scattered fairly freely throughout much of the image plane, they also show a tendency to be located close to bright emission rims and frontal structures. Where the rims arise from shocking with the SN outflow, then it is possible that the formation of the YSOs was triggered by the G357.7+0.3 event.

We have identified some 18 possible triggered YSOs, provisionally identified as Class I sources using the procedures outlined above. These are identified in Fig. 3 using the larger red disks. We have subsequently determined the fluxes of these sources at 24 μm (see Sect. 2), and determined the positions of the stars in the [3.6]-[5.8]/[8.0]-[24] colour plane (Fig. 4). The positions of the differing stellar regimes are based on the analyses of Reach et al. (2004) and Rho et al. (2006). It is clear that within the uncertainties associated with the YSO regimes, and errors deriving from the MIR photometry, only nine of the sources are potentially Class I YSOs. These are indicated in the figure by the filled disks. Although this might be thought, at first sight, to represent a rather paltry total of sources, the positional and colour constraints are extremely severe – much more so than in many other studies of the kind – and rates of star formation are relatively modest. Details of the positions of these stars are provided in Table 1, where we also include the IRAC and MIPSGAL photometric results. Photometric errors are indicated in parenthesis, whilst magnitudes at 24 μm are determined using the Vega calibration of Rho et al. (2006). Finally, upper limit MIPSGAL values correspond to $3\sigma_{BK}$ values, where $\sigma_{BK}$ is the standard deviation of the background fluctuations, normally dominated (for these stars) by variations in the emission of the surrounding clouds. The SEDs of the stars are later modelled in Sect. 6, where we will constrain a broad range of their physical parameters.

It is finally possible that several of the 24 μm sources not considered here may also represent examples of very young YSOs, and locally triggered star formation. However, most MIPSGAL sources having reasonably complete SEDS are likely to have been detected using the procedures outlined above. Those cases with inadequate IRAC or 2MASS photometry are of little utility in modelling the YSOs.

## 4. The MIR Structures of Five wind-Swept Clouds

It is clear that a large fraction of the clouds near G357.7+0.3 have wind-swept appearances, aligned tails and morphologies, and south-easterly pointing bright emission rims. It is likely that all of these characteristics were formed through a single event, and we shall be assuming that SNR G357.7+0.3 is the probable agent of change.

We consider the MIR morphologies and properties of five of these clouds in the analysis presented below.

### Region A

Images of source A at 5.8, 8.0 and 24.0 μm are illustrated in Fig. 5, where the results derive from the GLIMPSE and MIPSGAL surveys. We also show a composite image of all of these results in the lower right-hand panel, where we have represented 5.8 μm emission as blue, 8.0 μm emission as green, and 24.0 μm fluxes as red. Note that whilst the 5.8 and 8.0 μm results have similar levels of spatial resolution (the FWHM of the point spread functions is of order 1.7-2 arcsec, as noted in Sect. 2), the image at 24 μm has a significantly lower resolution (FWHM ~ 6 arcsec). This results in the red halos seen about the stars within this field.

It is apparent, from these results, that emission increases strongly to longer wavelengths. It is also clear that the region of interaction with the SNR leads to a bright emission edge, whose L-shaped morphology implies a wind direction from the lower left-hand side. Only two Class I YSOs are detected within the field, both of which are located within the primary cloud structure, and close to the bright emission rim. This positional correlation between the rims and YSOs will be seen as characteristic of most of the sources in this study. It is also possible that other Class I stars may occur within this region, and several sources were detected having [5.8]-[8.0] indices typical of less evolved YSOs. These sources have not been detected at 3.6 and/or 4.5 µm, however, and this precludes their identification within the diagnostic colour plane of Allen et al. (2004).

The blue circles in the lower right-hand panel indicate the YSOs used in our analysis in Sect. 6. These sources represent possible examples of triggered star formation. Similarly, we have indicated the position of the FIR point source IRAS17328-3050 through use of a darker blue (error uncertainty) ellipse. It would seem that this object is not related with any of the Class I sources, and may represent a Class 0 YSO – a source for which most of the IR flux arises from the accreting envelope, and for which MIR flux levels are relatively low.

Finally, it is worth noting the small regions of obscuration just above the right-hand YSO, and to the right of the cloud (close to $l$ = 357.236°, $b$ = 0.713°). These are most clearly visible in the 8.0 µm image, and may represent local regions of enhanced extinction. We shall see clearer evidence for such features in the other regions to be investigated here, and attempt to evaluate the levels of extinction associated with the clouds (see Sect. 5). Similarly, it should be noted that whilst the main frontal regime has a yellow appearance, testifying to appreciable emission in the 5.8 and 8.0 µm channels, the redder 24 µm emission is stronger in fainter regions of this source.

A profile through the leading edge of the cloud is illustrated in Fig. 6, where it will be seen that the ridge is detected at all of the IRAC wavelengths, from 3.6 µm through to 8.0 µm, and has a double-humped shape extending over ~14 arcsec. It is likely that each of the peaks correspond to individual frontal structures, whilst the extension of the profiles reflects complexity in the region, likely arising from distortion of the fronts by Rayleigh-Taylor and Kelvin-Helmholtz instabilities, and variations in the density of the impacted cloud.

Finally, we have determined the $F(3.6\mu m)/F(5.8\mu m)$ and $F(4.5\mu m)/F(8.0\mu m)$ ratios as described in Sect. 2, based upon relative surface brightnesses for the central portions of the profiles; that is, for the region -10 < RP < 3 arcsec corresponding to the primary frontal structures. The results have been plotted in the diagnostic colour plane of Reach et al. (2006) in Fig. 7, whence it is clear that the results fall within the regime expected for shock excited $H_2$ emission. Note that the CO fundamental band is also strong in Herbig-Haro shocks, and would be expected to increase 4.5 $\mu m$ fluxes, and shift sources to the right (Reach et al. 2006). The fact that the two regions (A & C) in Fig. 7 are located to the extreme left of the molecular regime suggests that this contribution is modest or non-existent.

At least two major biases may affect these values. On the one hand, it is possible that estimates of background emission may be somewhat in error, whilst on the other, contamination by field stars may increase both of the shorter wave fluxes. Although the sizes of these errors are difficult to determine, it is likely that each of them will tend to have opposing effects – errors in background are most likely to reduce the ratios, whilst contamination by stars will tend to increase them. Where this is not the case, however, and field star fluxes are larger than appears likely to be the case, then it is possible that the emission derives from polycyclic aromatic hydrocarbons (PAHs). This may arise due to an increase in the number densities of smaller (PAH carrying) grains, and/or in the levels of PAH band enhancement by local excitation processes. In the former case, compression of gas through local shock activity will be likely increase the local column densities of IS grains, although it is also possible that sputtering and vaporization will destroy the smaller dust particles (e.g. Allain et al. 1996; Jones et al. 1996). This latter problem, where it occurs, may be compensated by the shattering of larger grains, leading to a replenishment in the numbers of smaller dust particles (see e.g. Jones et al. 1996).

## 4.2 Region B

Images of this source between 5.8 and 24.0 $\mu m$ are illustrated in Fig. 8, where we have again produced a composite image in the lower right-hand panel. This structure is actually composed of two main components – a reddish plume centred on the Mira V1139 Sco ($l$ = 357.136°, $b$ = 0.840°), which is clearly dominated by the 24 μm emission, and a yellow convex rim located part of the way along the tail, which is strong in shorter wavelength emission, and may or may not be associated with the plume.

The nature of the plume is somewhat ambiguous, as mentioned in the introduction. Whilst the location of V1139 Sco may imply that it results from ram-pressure stripping of the Mira mass-loss envelope, we also note the presence of a couple of YSOs close to the head of the structure. It is therefore possible that we are witnessing the almost complete dissolution of the original ISM globule, leaving behind two or more of the YSOs created through interaction with the SNR wind. Were this to be the case, then the Mira would represent little more than a fortuitously placed foreground source. Whilst the distance to the Mira appears to differ from that of the SNR (see Phillips et al. 2009), the uncertainties in these estimates preclude us from deciding between these two options.

### 4.3 Region C

Region C is illustrated in Fig. 9, and represents one of the most interesting of this present batch of wind-fragmented globules. The cloud shows evidence for eight Class I YSOs close to the interaction region with the SNR, and a region of high extinction material close to $l$ = 357.084°, $b$ = +0.552°, most clearly seen in the 8.0 μm image. The positions of the YSOs, and their concentration about the frontal portions of the cloud, makes it very likely that the stars and cloud are intimately related, and that star formation was triggered by interaction with the SNR or pre-cursor wind.

Finally, the FIR sources IRAS 17329-3107 and IRAS 17330-3105 are indicated using blue error ellipses. It would again appear that there is little relation between these latter sources and the Class I YSOs.

A profile through the region is illustrated in Fig. 10, where it is clear that the high extinction region descends below background emission levels, whilst the primary peak (centred at relative position RP = 0 arcsec) is relatively narrow, having a FWHM of ~9.0 arcsec.

The fact that emission levels in the high extinction region fall below those of the background suggests that this, and the other clouds in this study, are closer than some of the more extended Galactic/SNR emission. Where this is the case, then one can make an attempt to evaluate levels of extinction, as described in our later discussion in Sect. 5. We determine that $A_V$ is likely to be of order ~23 mag.

The flux ratios for the primary peak in Fig. 10 (-7.8 arcsec < RP < 6.6 arcsec) place the source within the regime of shock excited $H_2$ emission, as illustrated in Fig. 7, and emission mechanisms are therefore likely to be very closely similar to those in Region A (see Sect. 4.1).

**4.4 Region D**

This source is illustrated in Fig. 11, and possesses an almost perfect convex frontal structure with changes in colour (lower right-hand panel) indicative of varying emission mechanisms. It is likely that PAH emission is important in the 5.8 and 8.0 $\mu$m channels, for instance, where the v = 0→0 S(4) to S(8) transitions of shock excited $H_2$ may also contribute to fluxes. Warm dust continua may also be important within these bands, although they are likely to be proportionately more important at 24 $\mu$m, where the v = 0→0 S(4) $\lambda$28.219 $\mu$m transition of $H_2$ also occurs. It is difficult, in the absence of spectroscopy, to define what the principal mechanisms might be, whilst the relative absence of shorter wave emission (see also our comments below) makes it difficult to determine the position of this source within the diagnostic colour plane (Fig. 7).

We note little evidence for localised, higher levels of extinction, or for associated star formation, and this, together with the low surface brightness of this feature, may imply that we are witnessing the rapid dissolution of a lower mass cloud. Profiles through the source are illustrated in Fig. 12, and show a bright narrow frontal peak at 8.0 $\mu$m, centred at RP = 0 arcsec, much weaker emission at 5.8 $\mu$m, and an absence of emission at shorter wavelengths.

**4.5 Region E**

Finally, the larger region E contains multiple evidence for bright rims, and likely shock and/or ionisation fronts, seen most clearly in the 24 μm image illustrated in Fig. 13. It also contains what are presumably highly extincted cloud cores close to $l$ = 356.518°, $b$ = 0.661°, and what appear to be weaker but similar extinction structures close to $l$ = 356.633°, $b$ = 0.798°. A blow-up of these regions at 8.0 and 24 μm is illustrated in Fig. 14, from which it is evident that most of these features have similar morphologies; the higher extinction regions appear to be extremely compact, and are located immediately behind the shock/ionisation fronts. Note that the effect of extinction is more obvious at 8.0 μm than it appears to be at 24 μm, where we see only the deepest portions of the obscuring clouds – a result of the decrease in extinction to longer wavelengths, as described in our analysis in Sect. 5. The usual sprinkling of YSOs occurs close to the presumed frontal structures, and we have again identified possible examples of triggered star formation in the lower right-hand panel (light-blue circles). The latter image also contains a single IRAS source at the extreme right, apparently associated with a bright MIR star.

Given the complexity of this region, and the extended nature of the emission, it would seem likely that a larger IS cloud is in the process of being destroyed, revealing regions of interaction with individual condensations within the cloud. Evidence for such clumps has previously been noted in other neutral regions, and discussions of this phenomenon can be found in Williams, Blitz & Stark (1995), Blitz (1993), Elmegreen & Falgarone (1996) and Kramer et al. (1998). It appears that clumps within molecular clouds are likely possess a broad range of masses, and a power-law spectrum dN(M) $\propto$ $M^{-\alpha}$dM; where dN is the number of clouds with mass between M and M+dM, and the exponent $\alpha$ is of order ~ 1.4-1.9. Most of the clumps are therefore small, with masses ranging down to ~$10^{-4}$ $M_{\odot}$. Where the shocked cloudlets in Fig. 2 represent the counterparts of such condensations, however, then it is very likely that we are observing higher mass components which have still not entirely been destroyed.

A profile through the high extinction core is illustrated in Fig. 15, which also includes a couple of the YSOs contained within this zone. A narrow rim is positioned close to RP = -15 arcsec, where it appears that a possible Class II YSO may also be located. This is subsequently followed by a deep decline towards RP = 0 arcsec, corresponding to the narrow absorption rim seen in Fig. 14, followed by a gradual recovery in emission levels to larger values of RP.

Our analysis of extinction for this dip (Sect. 5) suggests mean values of order $A_V \sim 26$ mag, whilst the YSOs associated with this and other frontal structures are considered in Sect. 6.

## 5. The High Extinction Nuclei in Regions C and E

We have noted, in Sect. 4, that several of the fragmented cloud structures possess extremely dark cores; regions in which emission declines appreciably below that of the local shocked structures, and more extended background component. Thus, we find that the dark nuclear regions of cloud E absorb up to ~73 % of 3.6 $\mu$m background emission, whilst the nuclear emission in cloud C declines to ~30 % of background. This suggests that the dark cores of these clouds contain appreciable quantities of dust, and that the background emission is located at greater distances than the clouds. Where this is not the case, and a proportion of the "background" arises at smaller distances from the Earth, then our estimates of extinction and cloud mass will correspond to lower limit values.

The analysis of extinction in these regions is open to a variety of uncertainties. Nevertheless, if we assume that areas adjacent to the cores have relatively low levels of extinction, then comparison with emission in the cores should permit reasonable estimates of $A_\lambda$. Note in this respect that foreground extinction is of relatively little importance, and is not considered in this analysis; we are only concerned with local extinction, and components of dust within the sources themselves.

This procedure has been employed to determine levels of extinction in regions C and E. The results are presented in Fig. 16, where the values for cloud C have been normalised to the 3.6 $\mu$m extinction of cloud E, and we have included estimates of 24 $\mu$m extinction deriving from the MIPSGAL project. Errors in these values are estimated by determining surface brightnesses at five locations about the cores, and evaluating the corresponding uncertainties in our estimates of $A_\lambda$.

We have finally fitted trends deriving from the empirical laws of Cardelli et al. (1989) and Martin & Whittet (1990); illustrated the least-squares power law fit $A_\lambda = 0.372(\log(\lambda/\mu m))^{-2.06}$; and included the MIR trends of Indebetouw et al. (2005), based upon their analysis of Spitzer IRAC data. It is clear from this that the MIR results of Indebetouw et al. (2005) bear little relation to what we appear to be observing here. Their suggested fall-off in $A_\lambda$ is very much smaller than is observed for the present clouds. On the other hand, the empirical trends of Cardelli et al. (1989) and Martin & Whittet (1990) fit the shorter wave trends reasonably well (i.e. the fall-off in extinctions where $\lambda < 8\,\mu m$), suggesting that the MIR variation can be represented using extrapolations from shorter IR wavelengths ($\leq 3.3\,\mu m$ in the case of Cardelli et al., and $< 5\,\mu m$ for Martin & Whittet). Having said this, however, there is some indication that the trends level out towards longer MIR wavelengths, and that the 24 $\mu m$ results are more elevated that would be expected from previous analyses.

We shall therefore the employ trends in $A_\lambda/A_V$ deduced by Cardelli et al. (1989), and the extinction values $A_\lambda$ arrived at here, to determine mean visual extinctions $A_V$ for the cores of Regions C and E. It is found that $<A_V> \cong 22.9$ mag for cloud C, and $<A_V> \cong 26.2$ mag for the darkest portion of cloud E (i.e. the narrow extinction region located immediately behind the bright emission rim; see Figs. 14 & 15). The broader region of lower extinction behind this front, evident in the gradual increase in profile strengths in Fig. 15, and the extended region of obscuration in Fig. 14, appears to have a typical mean value $<A_V> \cong 16.6$ mag. We designate the compact high extinction region in cloud E as "extinction region 1" (ER1), and the broader region of lower extinction as "extinction region 2" (ER2).

The extinction core of cloud C has a roughly spheroidal morphology, and we shall adopt a harmonic mean dimension for this region of $\theta_{HARM} = (\theta_{MAX}\theta_{MIN})^{0.5}$; where $\theta_{MAX}$ and $\theta_{MIN}$ are the maximum and minimum angular dimensions. The case of region E appears to be a little more complex however.

It is likely, for the latter region, that the cloud is being impacted at the lower left-hand side. This leads to the shocked higher extinction region ER1, located to one side of the original cloud ER2. If this interpretation is correct, then region ER1 may represent a disk-like regime oriented almost edge-on to the line of sight, with depth comparable to $\theta_{MAX}$.

Under these circumstances, a more appropriate measure of harmonic mean dimensions is given through $\theta_{HARM} = (1.5\theta_{MAX}^2\theta_{MIN})^{0.33}$ (i.e. the angular diameter of a sphere having the same physical volume). By contrast, the value of $\theta_{HARM}$ for ER2 is assumed to be given by $(\theta_{MAX}\theta_{MIN})^{0.5}$.

Finally, where one assumes that HI column densities are related to extinction $A_V$ through the relation $N(HI)/A_V \cong 1.5 \, 10^{21}$ cm$^{-2}$ mag$^{-1}$ (Diplas & Savage 1994), and that distances are of order ~4.0 kpc (Fontani et al. 2005; Phillips et al. 2009), then one can derive the atomic densities and masses indicated in Table 2.

Note however that the high extinctions found above suggest that the clouds are likely to have a large content of molecular material (levels of UV penetration would be too small to permit dissociation of the gas), implying corresponding Jeans masses of order $M_J \sim 730(T/K)^{1.5}(n(H_2)/cm^3)^{-0.5}$ $M_\odot$. Given the cloud masses and densities quoted in Table 2, then this implies that they are likely to be stable against collapse providing that temperatures are moderately high (> 50 K), and where there is an absence of external shocks. Such temperatures are rather larger than would be expected for isolated high extinction/density clouds, and this might imply that they have embedded star formation, leading to internal heating of the structures; that magnetic and turbulent virial terms are appreciable, and help to stabilise the clouds; or (and this is probably the least likely of all), that the clouds are unstable, and in the process of undergoing collapse. Whatever the situation, and despite these uncertainties, it is clear that they are sufficiently massive to represent current or future sites of star formation.

## 6. The Properties of the Triggered Class I YSOs

We have identified nine Class I YSOs which are likely to have been triggered through the interaction of SNR G357.7+0.3 with the ISM (see Sect. 3). Details of these stars are provided in Table 1. Where this is the case, then their lifetimes will be no greater than that of the SNR, which is determined to be of order $\sim 10^4(D/5 \text{ kpc})^2$ yr (Leahy 1989). Given a distance to the source of ~4.0 kpc (see Sect. 5), then this implies an overall age of $\leq 6.4 \, 10^3$ yrs. It is also possible that the stars were forming in pre-existing globules long before the supernova event occurred, however, and that interaction with the SN outflows has swept away the larger, lower density, and non-star-forming regions of IS cloud. There is indeed some evidence for this process in region E, where elements of a larger cloud structure still appear to be in place. This process of "revealed" star formation cannot entirely be discounted, and represents a caveat to the analysis we shall be undertaking below. Where this occurs, then our constraints upon YSO lifetimes may be in error. Nevertheless, we believe that the locations of the sources so close to the shocked rims is likely to testify to a more dramatic origin for the YSOs, and imply abbreviated evolutionary lifetimes.

Given this situation, and assuming distances $3 \leq D \leq 5$ kpc, and levels of extinction $0 \leq A_V \leq 50$ (this is largely unknown, and we therefore place the widest plausible limits upon this parameter), then one can use the 2D radiative transfer modelling of Robitaille et al. (2006) to undertake a fitting of the individual YSO SEDs. The SEDs are determined using 2MASS, GLIMPSE, MSX, and 24 $\mu$m MIPSGAL photometry (see Sects. 2 & 3), and an example one of these SEDs is illustrated in Fig. 17, where we show model fits for the star at $l = 357.076°$, $b = +0.544°$ (upper left-hand panel). The grey curves represent various best fit models, whilst the dashed curve is the IS reddened stellar continuum, excluding the effects of local circumstellar dust extinction. The upper right-hand panel, by contrast, shows differing elements of emission for the best model fit, where the total flux is indicated by the black curve, the stellar photospheric flux is indicated by the dashed line (this is the flux prior to reddening by circumstellar dust); the disk flux is green; the scattered flux is yellow; the envelope flux is red; and the thermal flux is orange. Unless otherwise stated, the results include the effects of circumstellar extinction, but not of IS extinction. They also assume a representative distance of 1 kpc, and an aperture of close to 5 arcsec.

Parameters for the best fit models are illustrated in the lower panels of Fig. 17 (indicated by the solid bullets), where we show the dependence of stellar mass, luminosity, temperature and radius upon the evolutionary age of the source $T_{EV}$. The grey regions denote the ranges of parameter which were explored in these solutions. Where $T_{EV}$ is < $6.4 \ 10^3$ yr, then this implies very narrow ranges of values for all of these parameters. Details of selected parameters for the all of the candidate sources are also summarised in Table 1 (columns 9-12), where values in parenthesis indicate the logarithmic ranges in the parameters.

Note that the sources are located well away from the centre of the SNR, whence their ages may be less than $T_{EV}$. It is also possible that that the sources were triggered by pre-cursor winds, however, for which case ages may turn out to be somewhat larger. In either case, uncertainties in $T_{EV}$ by factors of ~2 have little impact on the modelling, or the final distributions of stellar parameters.

Generally speaking, where one now defines a range of model masses $\Delta M_i$ for all of the individual YSOs *i*, and plots these in a histogram with height $\Delta M_i^{-1}$, centred upon the mean masses $<M_i>$, then one can determine the distribution of masses for the complete sample of YSOs. Such a procedure ensures that all of the YSOs are given equal weighting within the graph.

Examples of such an analysis for the stellar masses and other parameters are illustrated in Fig. 18, where the histograms have in all cases been normalised to unity. Although the sample is too small to gain a reliable impression of triggered star formation as a whole – and is any case subject to a variety of biases, including those related to YSO evolutionary rates and luminosities, the sample size and limiting SST magnitudes (see e.g. Marquez-Lugo & Phillips 2010) - it does give a global sense of the nature of these YSOs. It is clear for instance that masses and luminosities cover the full ranges of value to be expected in star-formation zones. Stellar radii extend between ~8 and $10^2$ $R_\odot$, envelope accretion rates appear to be reasonably high ($3 \times 10^{-6}$ < $(dM/dt)_{ENV}/M_\odot yr^{-1}$ < $2 \times 10^{-3}$; a result of preferentially selecting Class I YSOs) whilst stellar photospheric temperatures are more restricted, and lie mostly in the range 4000 < $T_{EFF}/K$ < 5000. Finally, it will be noted that disk masses are reasonably low (typically ~ $10^{-3}$ -$10^{-1}$ $M_\odot$), and that most disk radii extend between ~ 2 and 25 AU.

Tighter constraints upon the characteristics of Class I sources will have to wait for deeper surveys, and larger population samples, whilst such an analysis could also be usefully expanded to include Class II YSOs as well.

## 7. Conclusions

We have undertaken an analysis of five regions of interaction between the SNR G357.7+0.3 and the ISM. The regions show evidence for cometary structures, bright rims, and the local formation of YSOs. It is noted that several of the regions also show evidence for high extinction cores, seen in silhouette against the shock emission interfaces, and the more extended background emission. An analysis of extinctions in these regions suggests that whilst the wavelength dependence of extinction is consistent with the trends of Martin & Whittet (1990) and Cardelli et al. (1989), it appears inconsistent with that of Indebetouw et al. (2005). We determine that mean extinctions $<A_V>$ are likely to range between 26 and 70 mag, cloud densities n(HI) are in the region of $\approx 10^4$ cm$^{-3}$, and that the masses of the clouds are of order ~40-106 $M_\odot$. It is therefore clear that these dense cloud nuclei have the potential to be future star formation sites. The westerly region E also shows evidence for having been a much larger cloud in the recent past, and appears to be in the process of being destroyed by shocks and/or ionisation fronts. It reveals evidence for the denser interior condensations which are probably typical of many IS clouds, but which are still too massive (in this present case) to have been destroyed by the SN winds.

The SNR/ISM interfaces appear to have flux ratios consistent with excited $H_2$ emission, although the possibility of strong PAH emission cannot entirely be excluded. The latter circumstance, where it occurs, may imply that grains are being shattered within the frontal structures.

Finally, we have determined the distribution of Class I sources within the field of windswept clouds, and identified those close to the bright emission rims as representing likely examples of SN triggered stars. After determining 24 $\mu$m MIPSGAL fluxes for these latter YSOs, and further filtering the sources in the MIPSGAL-IRAC colour plane, we have determined the SEDs of the stars using a variety of data, including 2MASS, IRAC, MIPSGAL and MSX photometry. These SEDs have subsequently been analysed using the 2D radiative transfer modelling of Robitaillle et al. (2006); a procedure which has permitted us to place conservative limits upon their luminosities, masses, temperatures and radii, as well as upon various characteristics of the circumstellar material. It would appear that the Class I YSOs have broad ranges of most of these parameters, including masses (M < 20 $M_\odot$) and luminosities (L< $10^4$ $L_\odot$), although temperatures appear to be relatively tightly constrained (4000 < $T_{EFF}$/ K < 5000).

**Acknowledgements**

We would like to thank an anonymous referee for his interesting remarks, several of which led to useful changes in the presentation of the paper. This work is based, in part, on observations made with the Spitzer Space Telescope, which is operated by the Jet Propulsion Laboratory, California Institute of Technology under a contract with NASA. Support for this work was provided by an award issued by JPL/Caltech. It also makes use of data products from the Two Micron All Sky Survey, which is a joint project of the University of Massachusetts and the Infrared Processing and Analysis Center/California Institute of Technology, funded by the National Aeronautics and Space Administration and the National Science Foundation.

Table 1

Spitzer MIR Photometry and Physical Parameters for
Nine Class I Triggered YSOs

| DESIGNATION | R.A. | DEC. | IRAC[1] | | | | MIPS[1] | Log(M/M$_\odot$)[2] ($\Delta$log(M/M$_\odot$)) | log(R/R$_\odot$)[2] ($\Delta$log(R/R$_\odot$)) | log(T/K)[2] ($\Delta$log(T/K)) | log(L/L$_\odot$)[2] ($\Delta$(log(L/L$_\odot$))) |
|---|---|---|---|---|---|---|---|---|---|---|---|
| | | | mag(3.6) | mag(4.5) | mag(5.8) | mag(8.0) | mag(24) | | | | |
| G357.2519+00.7168 | 264.02371 | -30.886971 | 14.84(0.17) | 13.07(0.10) | 12.00(0.11) | 11.58(0.16) | >6.60 | 0.06(2.11) | 1.30(1.40) | 3.59(0.22) | 1.80(3.60) |
| G357.1416+00.8406 | 263.832883 | -30.913017 | 9.09(0.07) | 8.14(0.05) | 7.35(0.03) | 6.59(0.03) | 4.00(0.01) | 0.04(0.08) | 1.04(0.08) | 3.60(0.01) | 2.51(0.07) |
| G357.0381+00.5719 | 264.031682 | -31.145171 | 12.69(0.05) | 11.74(0.07) | 10.76(0.06) | 10.22(0.10) | 7.10(0.17) | -0.17(0.26) | 0.95(0.10) | 3.59(0.02) | 1.24(0.50) |
| G357.0758+00.5440 | 264.082898 | -31.128507 | 10.50(0.04) | 9.64(0.05) | 8.71(0.03) | 7.05(0.03) | 3.19(0.01) | 0.11(0.30) | 1.08(0.25) | 3.61(0.04) | 1.57(0.54) |
| G357.0837+00.5431 | 264.088768 | -31.122267 | 10.75(0.09) | 9.40(0.06) | 8.29(0.03) | 7.45(0.03) | 4.36(0.03) | 0.24(0.48) | 1.22(0.28) | 3.62(0.03) | 2.36(0.49) |
| G357.0853+00.5528 | 264.080164 | -31.115734 | 13.75(0.08) | 11.91(0.08) | 10.94(0.07) | 10.41(0.05) | >5.09 | 1.00(0.73) | 1.73(0.70) | 3.65(0.10) | 2.65(0.70) |
| G357.0882+00.5514 | 264.083342 | -31.114035 | 9.07(0.03) | 7.87(0.04) | 6.99(0.02) | 6.21(0.03) | 3.18(0.01) | 0.50(0.55) | 1.37(0.51) | 3.61(0.02) | 2.61(0.35) |
| G356.5162+00.6664 | 263.608343 | -31.533154 | 11.85(0.06) | 11.03(0.05) | 10.36(0.05) | 9.70(0.04) | 6.93(0.16) | -0.11(0.22) | 0.96(0.23) | 3.59(0.03) | 1.24(0.48) |
| G356.5269+00.7173 | 263.565019 | -31.496603 | 10.79(0.04) | 10.04(0.04) | 9.48(0.04) | 8.96(0.03) | >5.76 | 1.10(0.04) | 2.04(0.08) | 3.65(0.03) | 3.45(0.30) |

1) Values in parenthesis correspond to the photometric errors; 2) Values in parenthesis correspond to the logarithmic ranges of the parameters.

Table 2

Properties of the High Extinction Cores in Regions C and E

| CLOUD REGION | $\theta_{HARM}$ | MEAN DIAMETER | <A$_V$> | n(HI) | MASS |
|---|---|---|---|---|---|
| | arcsec | pc | mag | cm$^{-3}$ | M$_\odot$ |
| C | 33.6 | 0.65 | 22.9 | 1.7 10$^4$ | 71.7 |
| E (ER1) | 37.4 | 0.73 | 26.2 | 8.1 10$^3$ | 40.0 |
| E (ER2) | 52.0 | 1.01 | 16.6 | 8.0 10$^3$ | 106.2 |

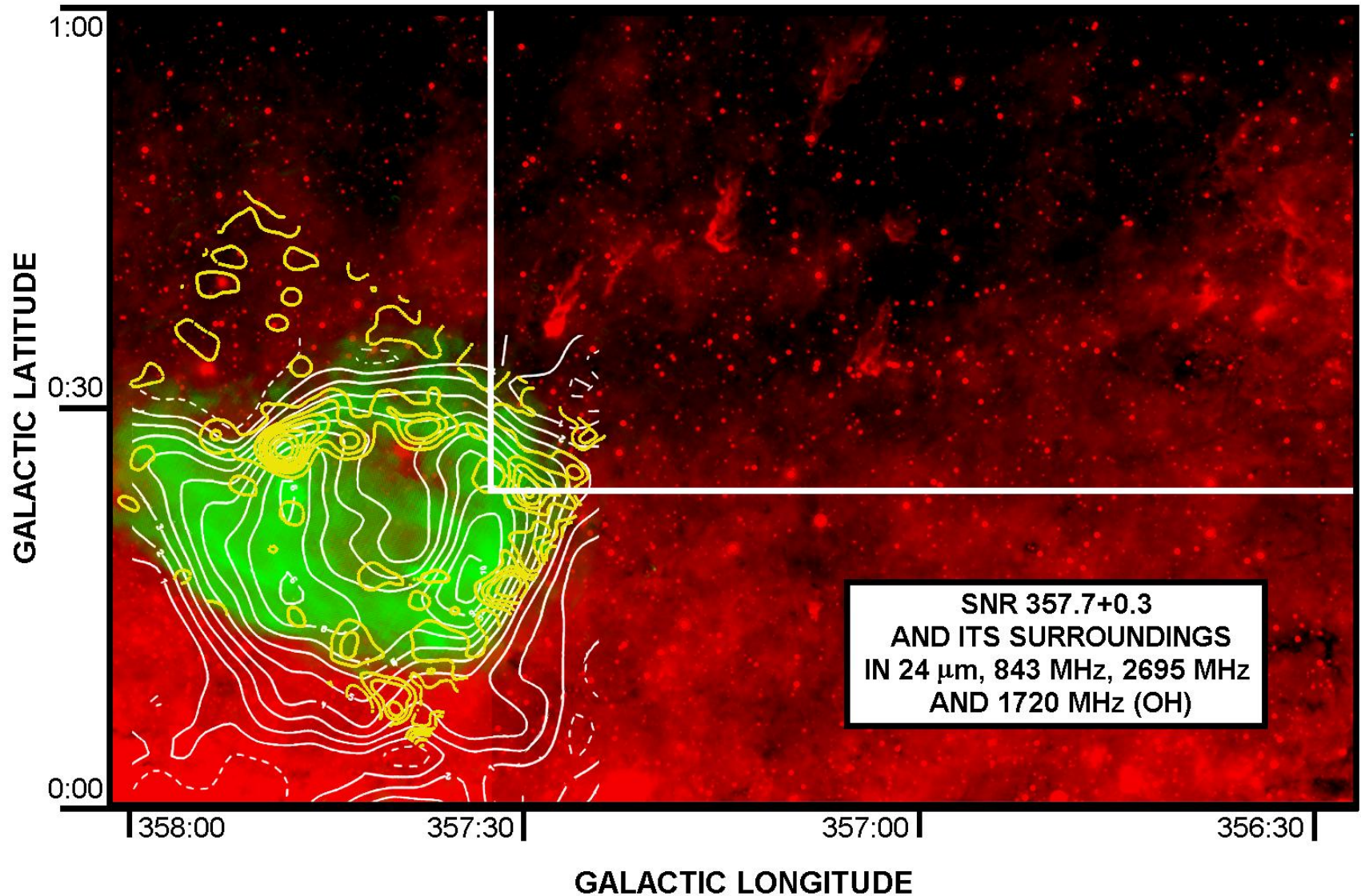

**Figure 1:** The interstellar medium in the vicinity of SNR G357.7+0.3, where we show a 24 μm mosaic deriving from the MIPSGAL survey (red), an 834 MHz image deriving from Gray (1994) (green emission); a 2695 MHz map obtained by Reich & Fürst (1984) (white contours); and 1720 MHz OH emission detected by Yusef-Zadeh et al. (1999) (yellow contours). The upper right hand portion of the image, identified using the white lines, appears to represent a region of small wind-swept clouds and bright emission rims, many of which are oriented towards the SNR.

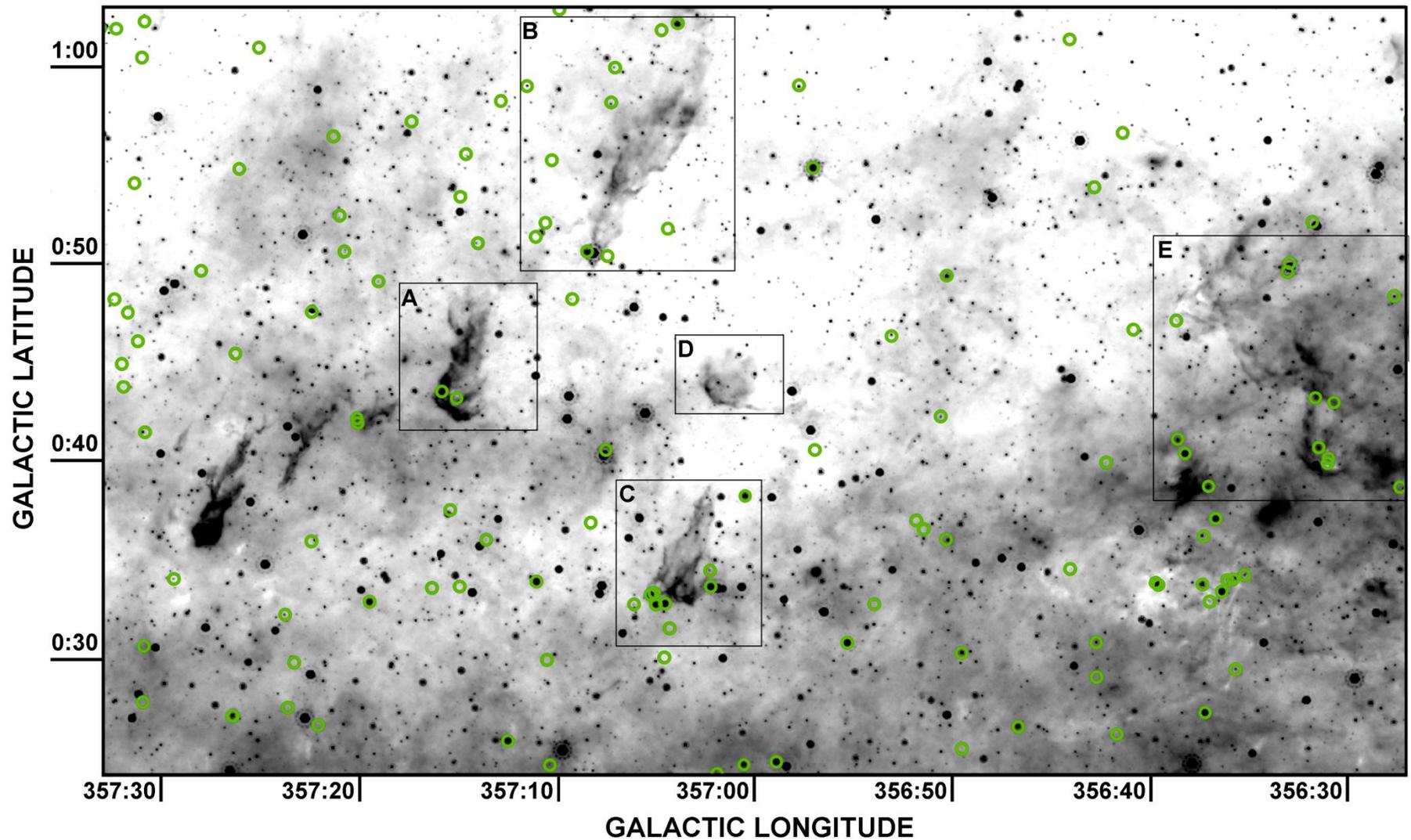

**Figure 2:** A 24 μm image of the region of interaction delineated in Fig. 1. The SNR is located to the lower left-hand side. The regions investigated in the present analysis are defined by the boxes, and labelled A through E. Similarly, the green circles indicate the locations of Class I YSOs, based upon the colours of the stars within the GLIMPSE II point source catalogue.

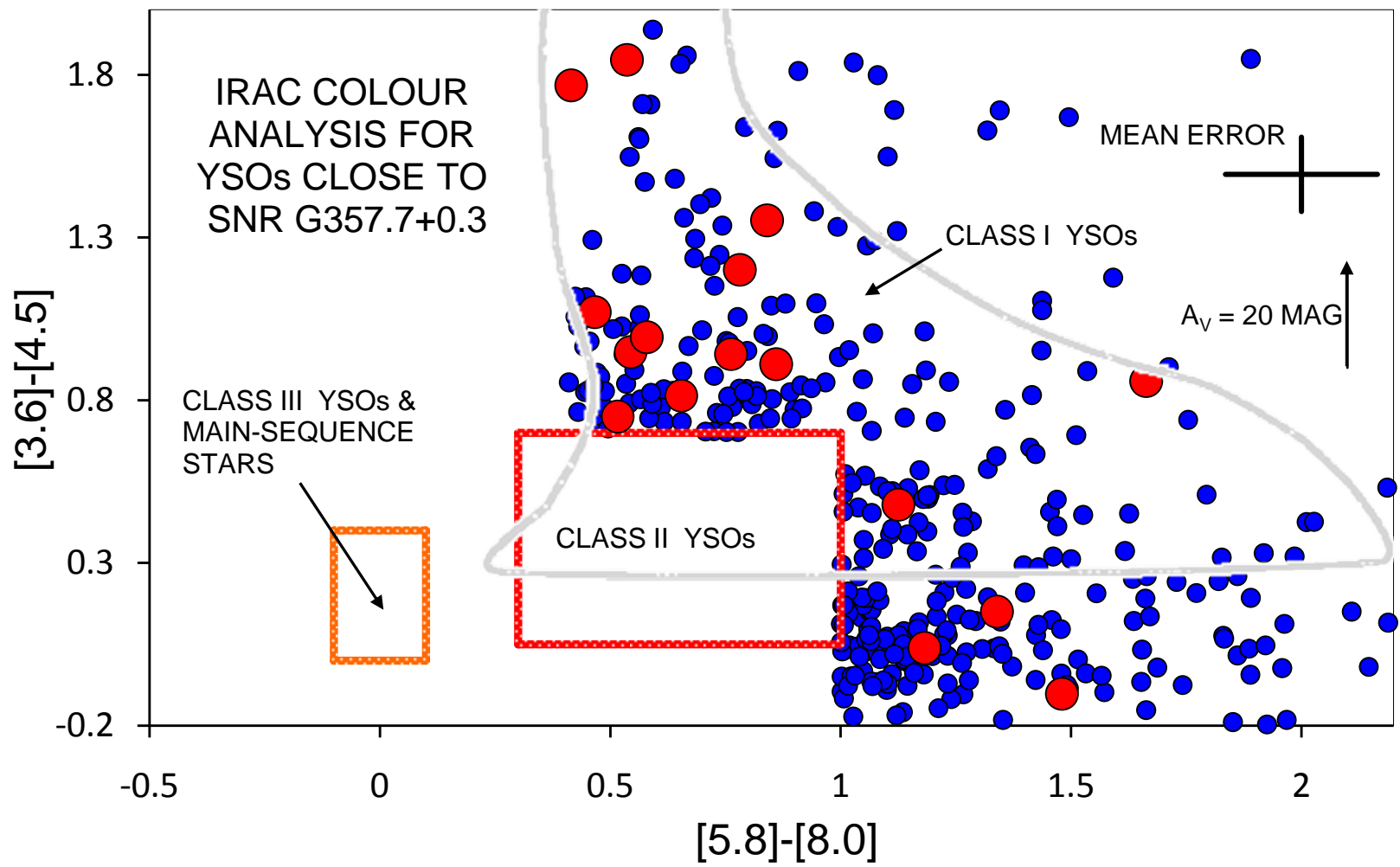

**Figure 3:** The positions of the YSOs in Fig. 2 (small blue symbols), and of possible triggered YSOs (large red symbols) in the MIR colour plane, where we also indicate the positions of various categories of YSO (based on the analysis of Allen et al. (2004)), the mean error in the colour estimates, and the reddening vector for $A_V = 20$ mag.

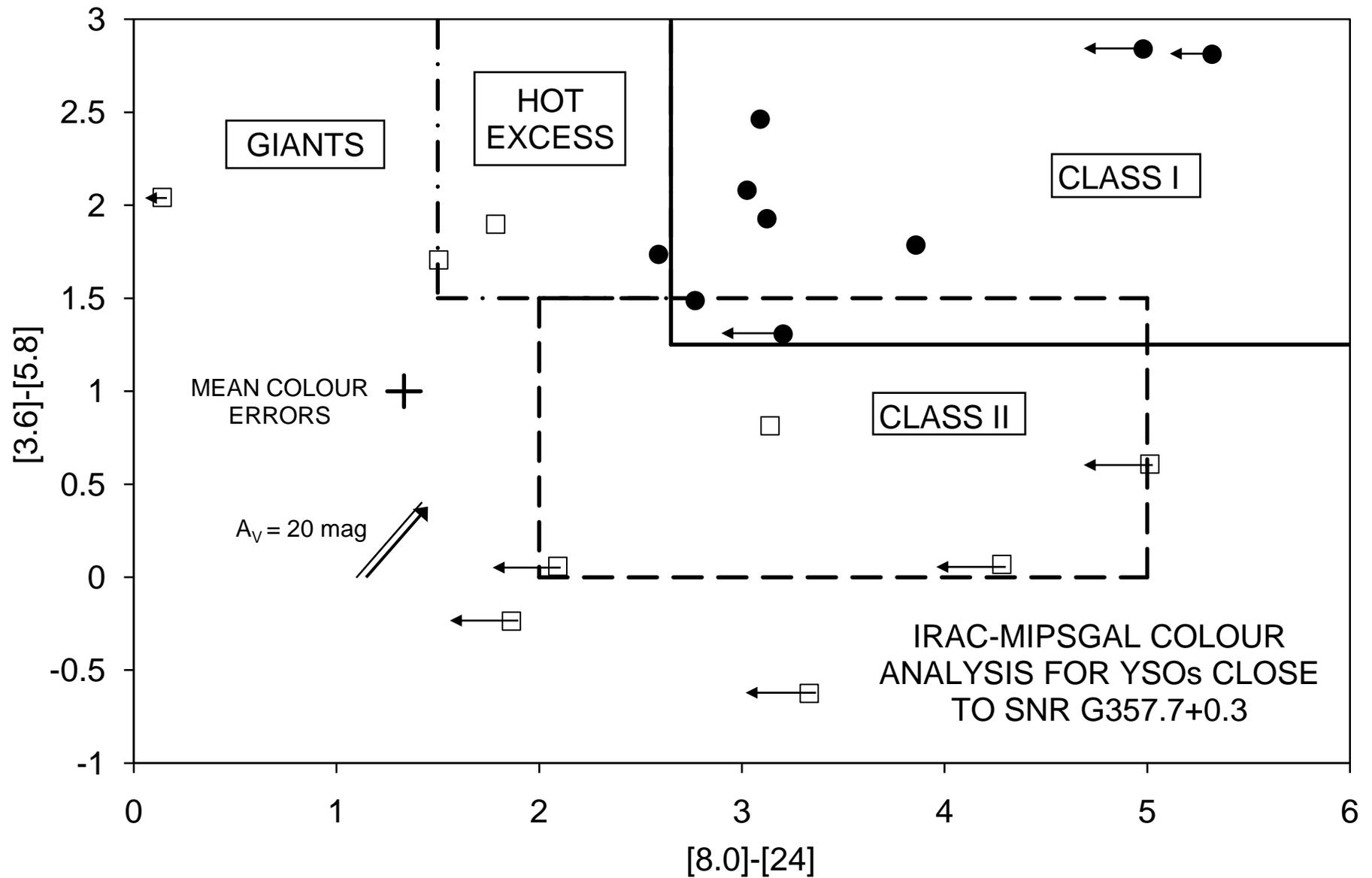

**Figure 4:** The positions of the candidate triggered YSOs in the IRAC-MIPSGAL colour plane, where we have also indicated approximate regimes of red giant and hot excess stars, and Class I & II YSOs. Upper limit colours are indicated using leftward pointing arrows, whilst the filled circles indicate likely Class I YSOs. The reddening vector is based on the analysis of Weingartner & Draine (2001), whilst the areas occupied by the differing classes of star are based on the analyses of Reach et al. (2004) and Rho et al. (2006).

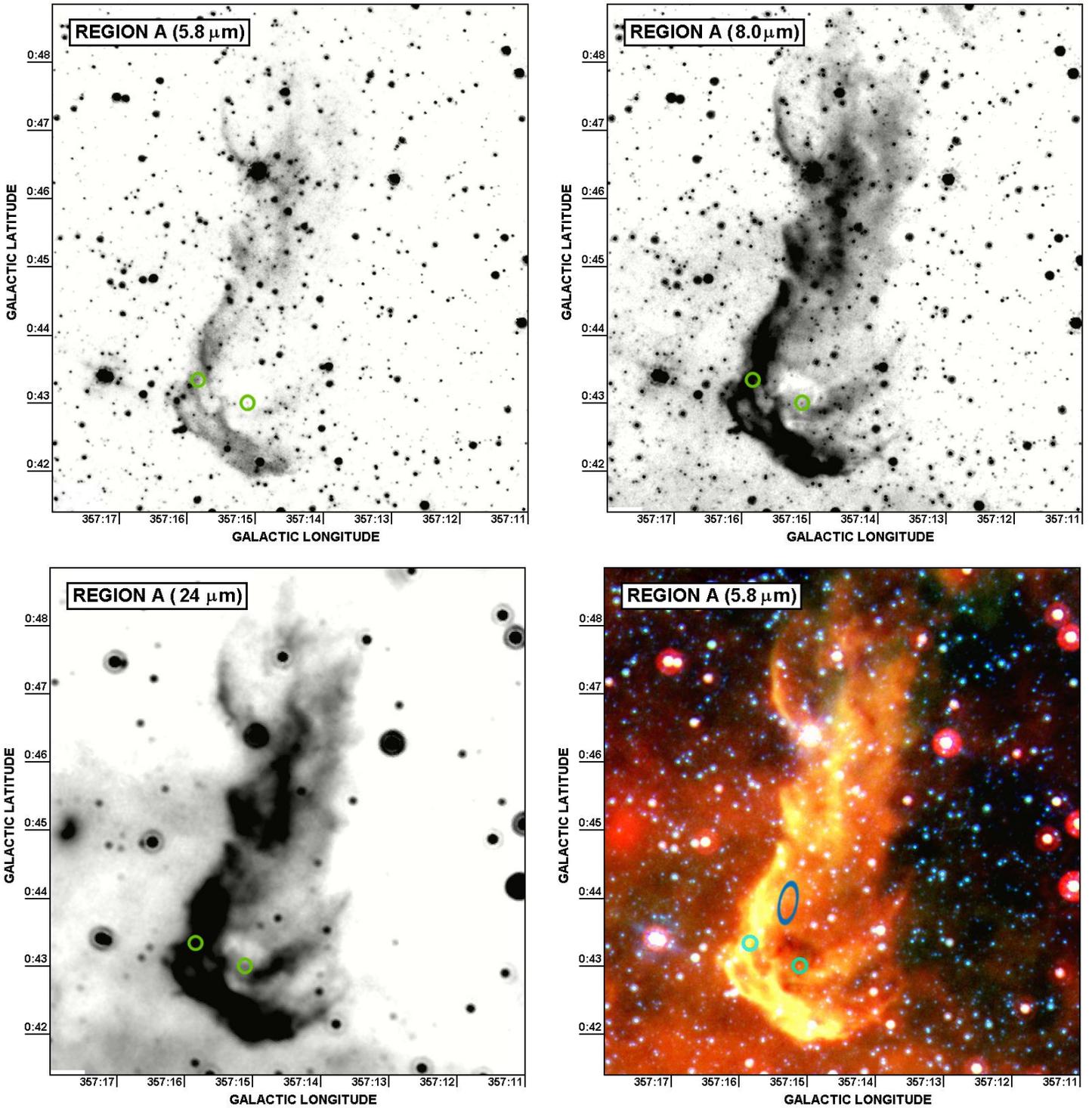

**Figure 5:** Cloud region A at 5.8, 8.0 and 24.0 μm, where the results derive from the GLIMPSE and MIPSGAL surveys. The lower right-hand panel is a composite of the three other panels, where 5.8 μm emission is represented as blue, 8.0 μm emission is green, and 24.0 μm fluxes are red. Green circles show the locations of Class I YSOs (they are represented as blue in the lower right-hand panel, to indicate their inclusion in the analysis of Sect. 6), whilst the darker blue ellipse corresponds to the error uncertainty associated with IRAS17328-3050.

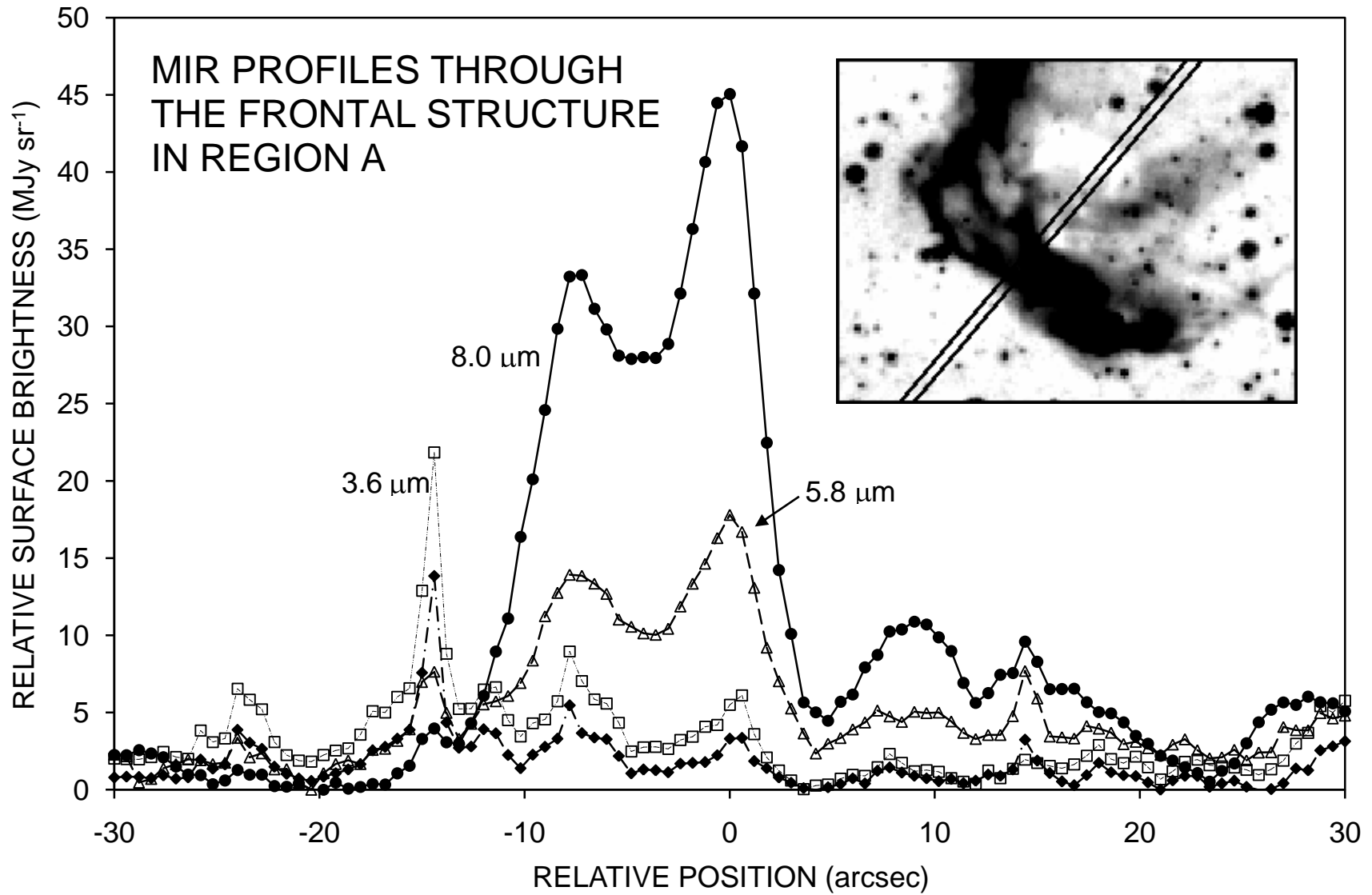

**Figure 6:** MIR profiles through the rim of region A, where the direction and widths of the slices are indicated in the inserted image. Each of the peaks is likely correspond to an individual shock front, although it is clear that the overall structure of the rim is complex, and affected by irregularities in the frontal structures. Both here, and in subsequent profiles, we represent the 8.0 μm trends with filled bullets, 5.8 μm emission with open triangles, 4.5 μm emission with filled diamonds, and 3.6 μm emission with open squares. The vertical axis corresponds to the relative surface brightness. Original surface brightnesses can be retrieved by adding the factors (Δ3.6, Δ4.5, Δ5.8, Δ8.0) = (3.22, 2.66, 11.27, 41.69) MJy sr$^{-1}$, as described in Sect. 2.

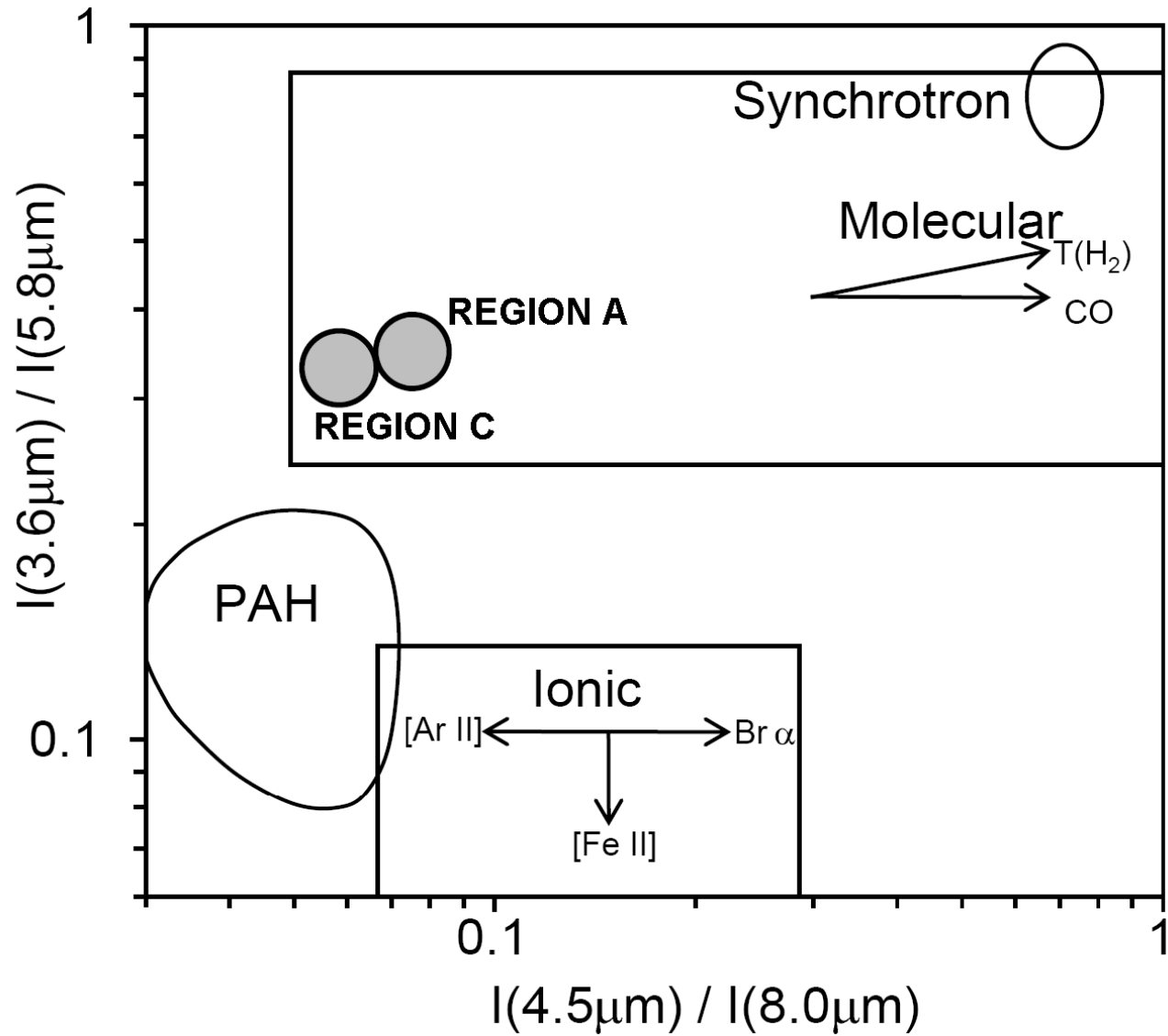

**Figure 7:** Positioning of the cloud frontal structures A & C within the diagnostic colour plane of Reach et al. (2006), whence it would appear that both of the sources show evidence for shock excited $H_2$ emission. Note however that uncertainties in the ratios may be sufficient to shift the sources into the regime defining PAH band emission.

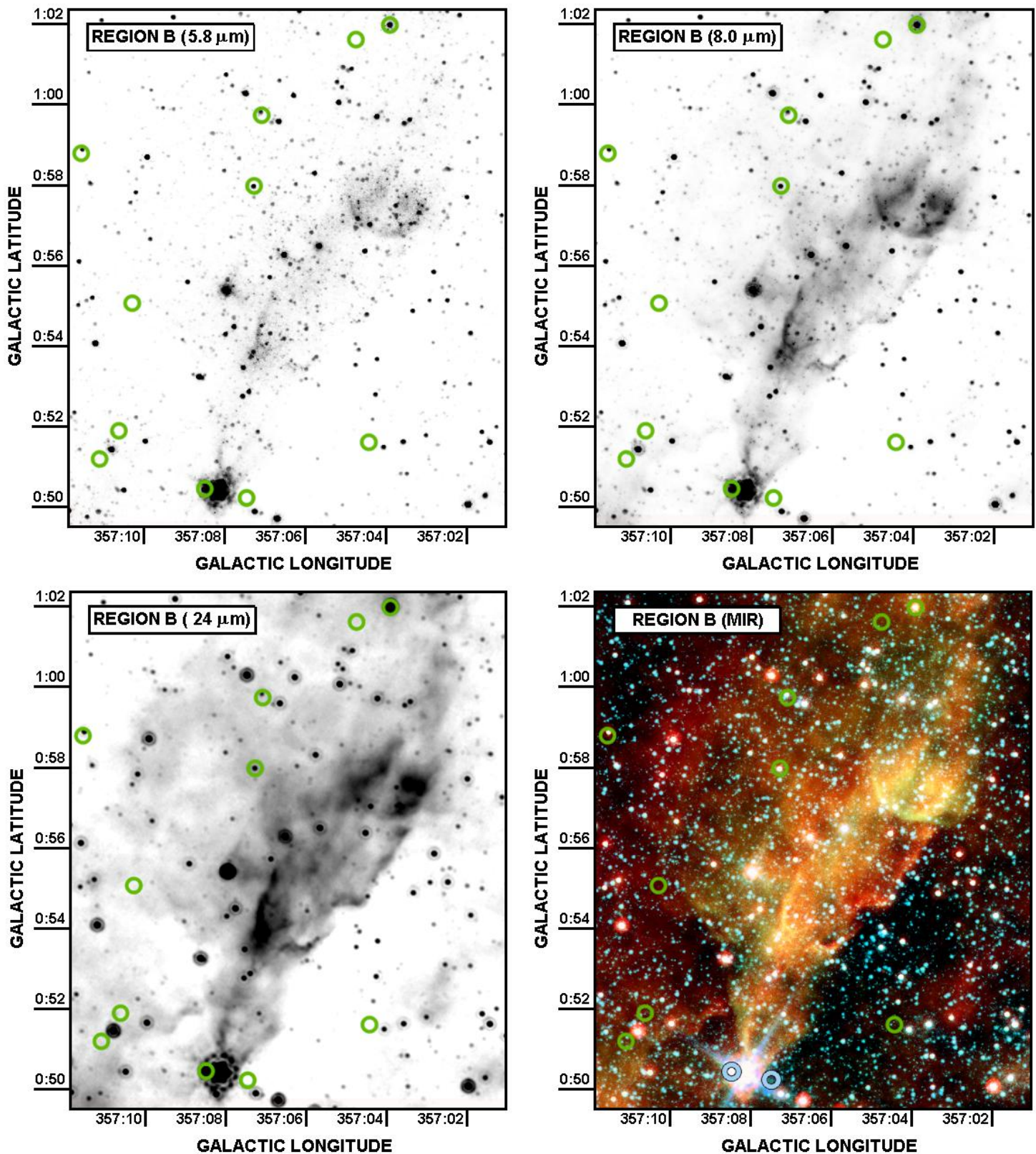

**Figure 8:** As for Fig. 5, but for region B. The bright star at the head of the plume is the Mira variable V1139 Sco.

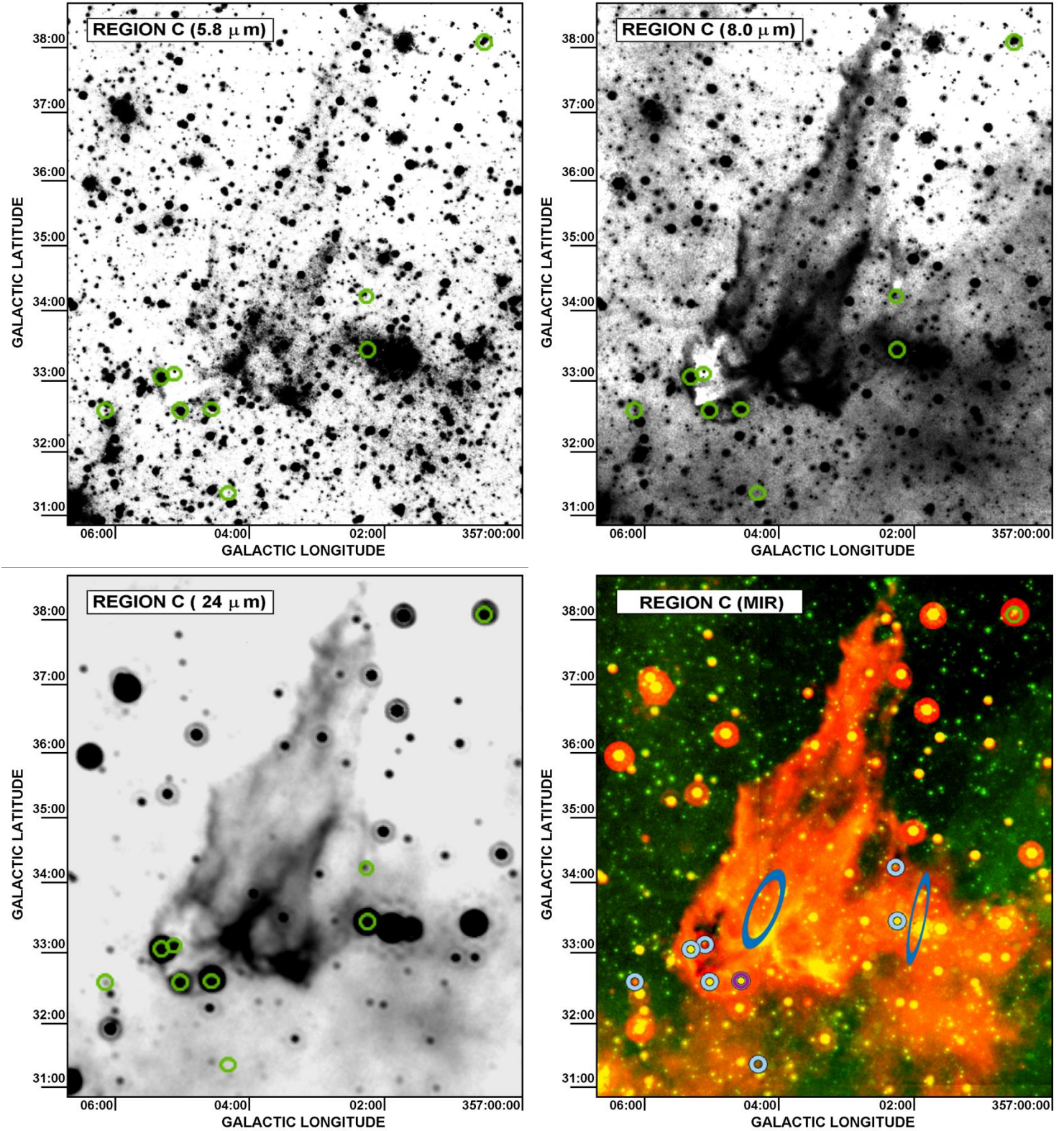

**Figure 9:** As for Fig. 5, but for the case of region C. Note the presence of multiple YSOs close to the rim of the structure, and of a dark extinction core to the lower left-hand side of the cloud (close to $l$ = 357.084°, $b$ = +0.552°); a feature which is seen most clearly in the 8.0 μm image.

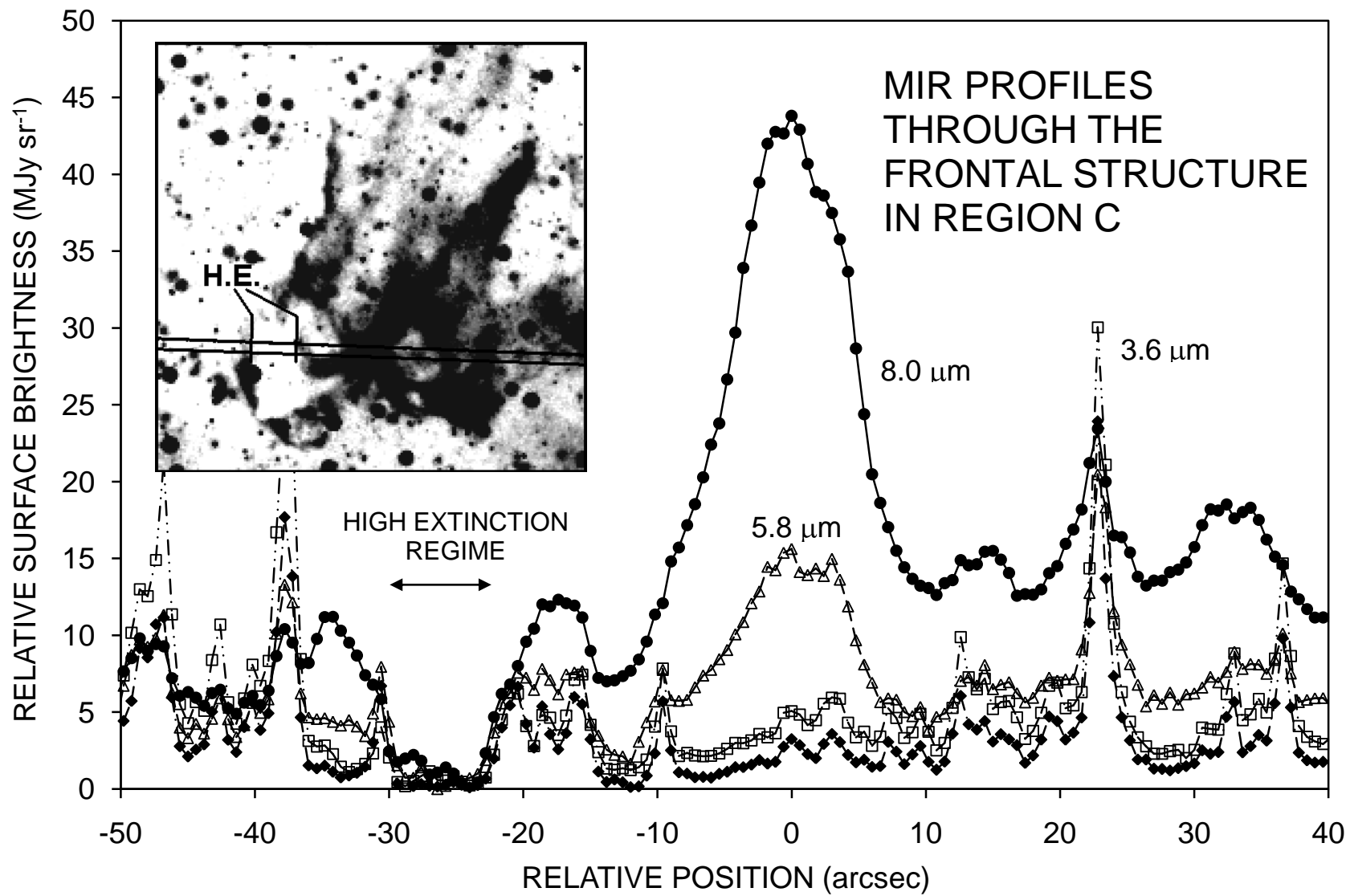

**Figure 10:** As for Fig. 6, but for a slice through region C. Note the deep dip between RP = -30→20 arcsec, likely caused by the silhouette of a high extinction core. This region is also indicated in the inserted image by the letters H.E., wherein we also show the direction and width of the MIR emission profiles. The primary emission peak shows evidence for fluxes at 5.8 and 8.0 μm, but relatively little emission at shorter MIR wavelengths. The profile correction parameters ($\Delta 3.6, \Delta 4.5, \Delta 5.8, \Delta 8.0$) are given by (2.27, 2.02, 11.43, 38.31) MJy sr$^{-1}$.

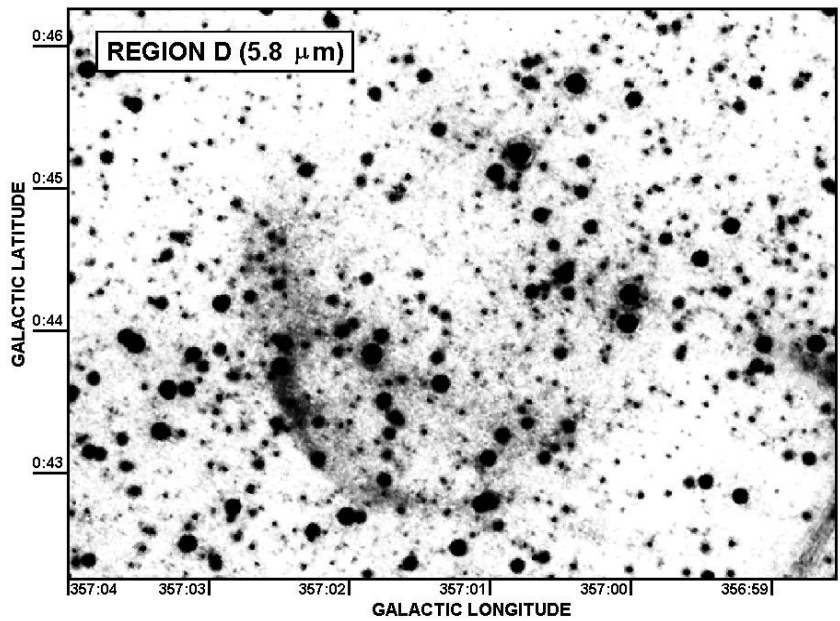 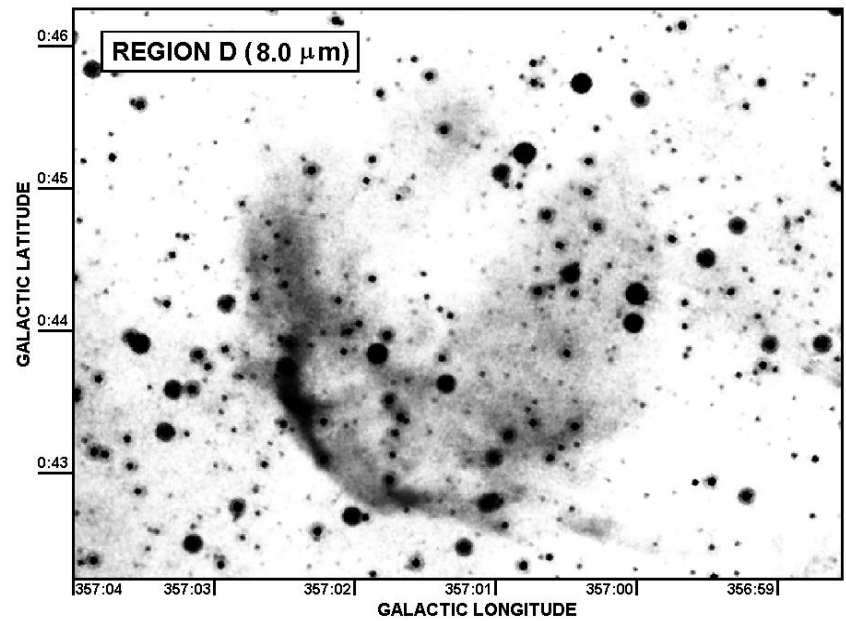
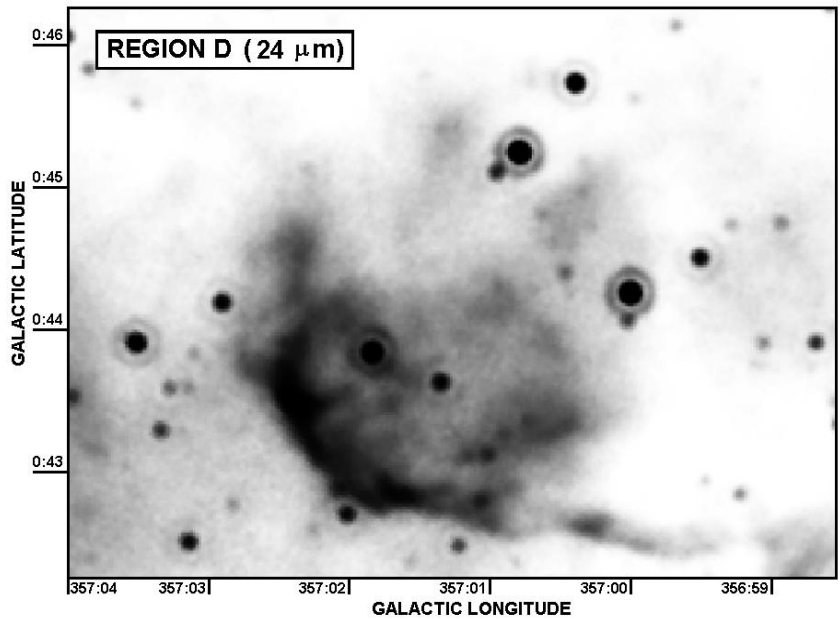 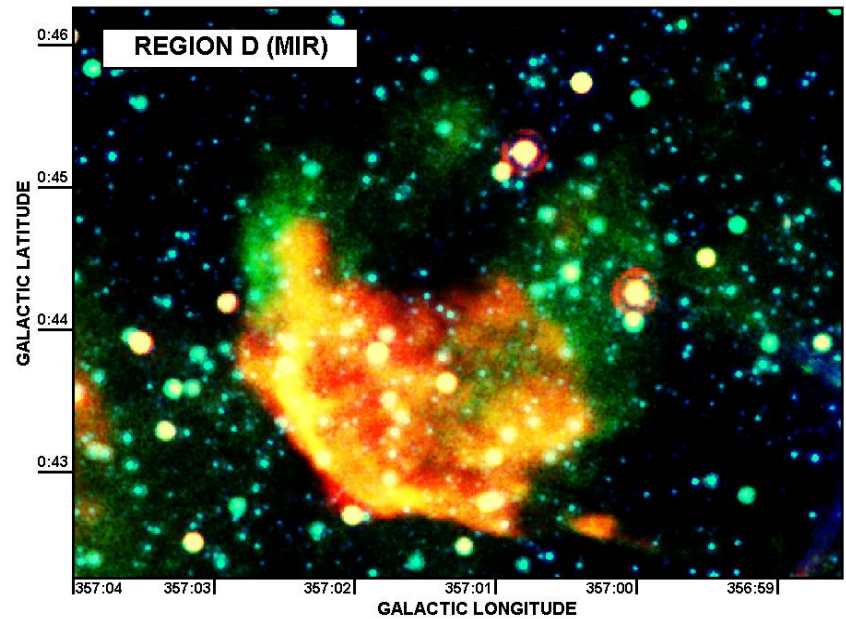

**Figure 11:** As for Fig. 5, but for region D. The lower right hand image shows particularly clear variations in colour, likely indicating a variety of emission mechanisms, whilst the cloud as a whole has a relatively simple convex morphology. The surface brightness of the structure is low, and there is no evidence for associated Class I YSOs.

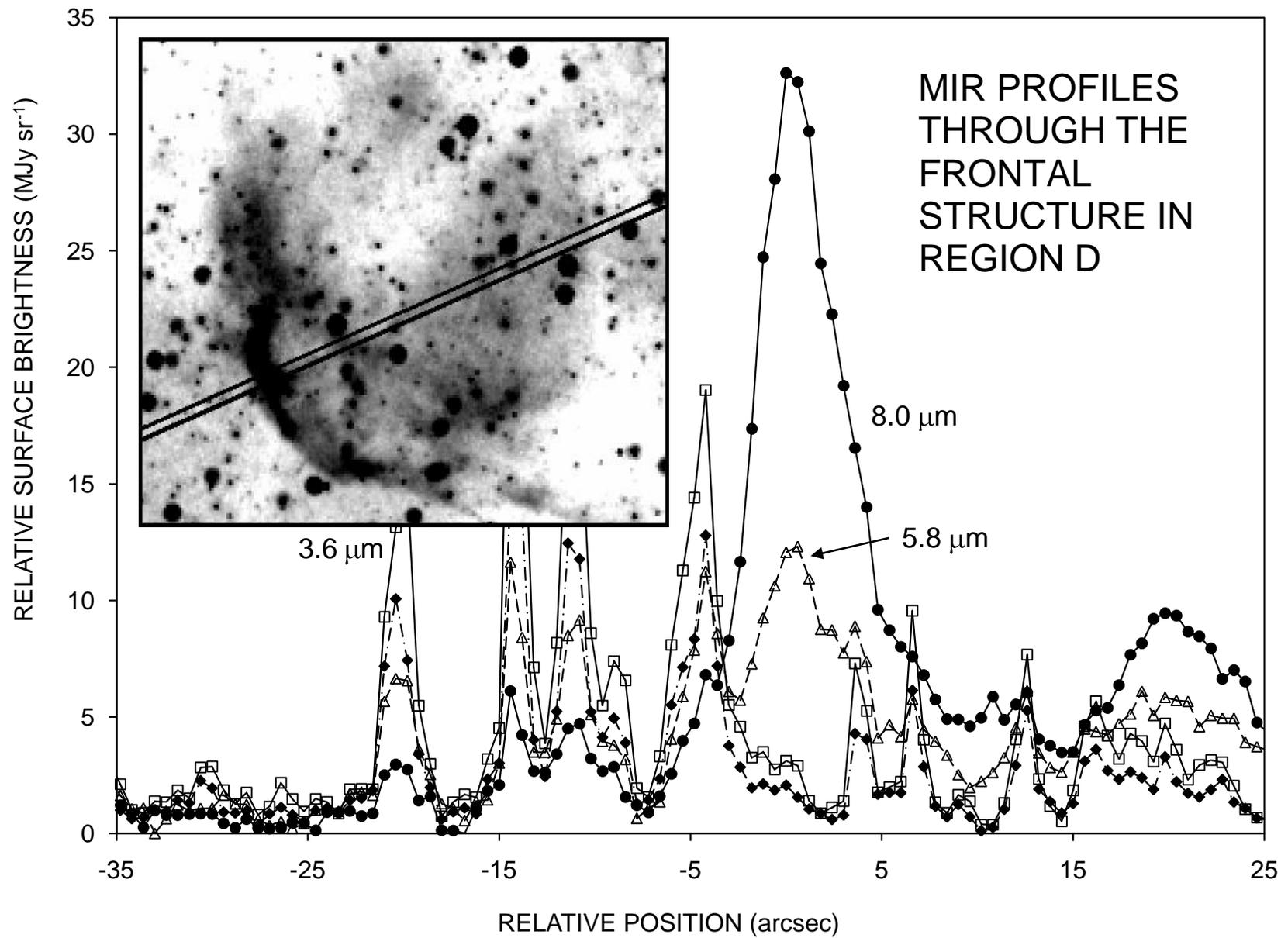

**Figure 12:** As for Fig. 6, but for region D. The 5.8 and 8.0 μm peaks at RP = 0 arcsec correspond to the narrow emission rim, whilst the profile parameters ($\Delta 3.6$, $\Delta 4.5$, $\Delta 5.8$, $\Delta 8.0$) are given by (4.41, 2.85, 11.147, 37.42) MJy sr$^{-1}$.

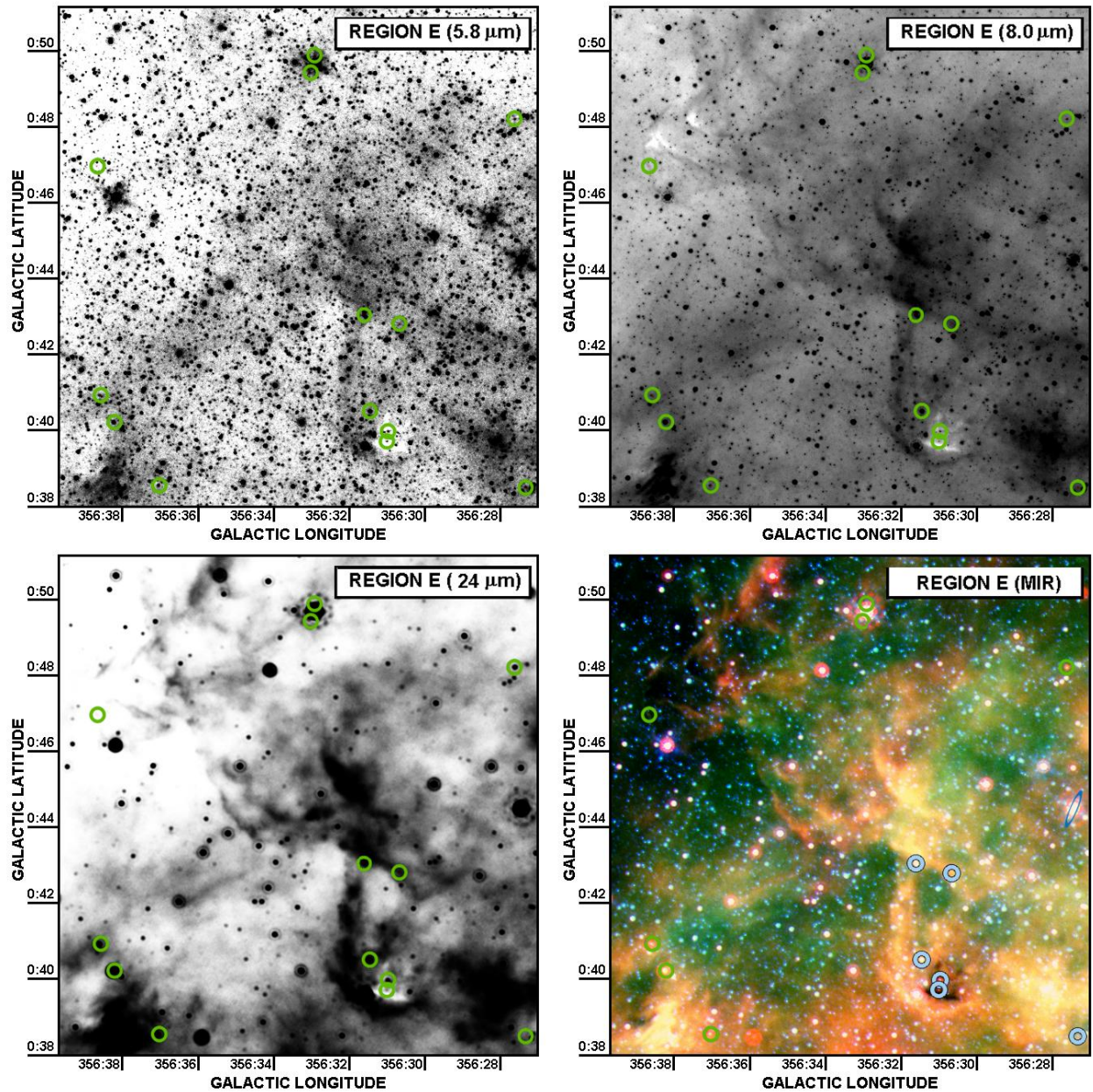

**Figure 13:** As for Fig. 5, but for the extended region of interaction E to the right-hand side of Fig. 2. Note the presence of multiple bright rims (particularly clear in the 24.0 μm image), suggesting interaction between the SN winds and individual cloudlets. It is likely that an extended IS cloud is at the point of dissolution, and that the shocking of this cloud is revealing internal density inhomogeneities. Note also the dark high extinction core to the lower right-hand side.

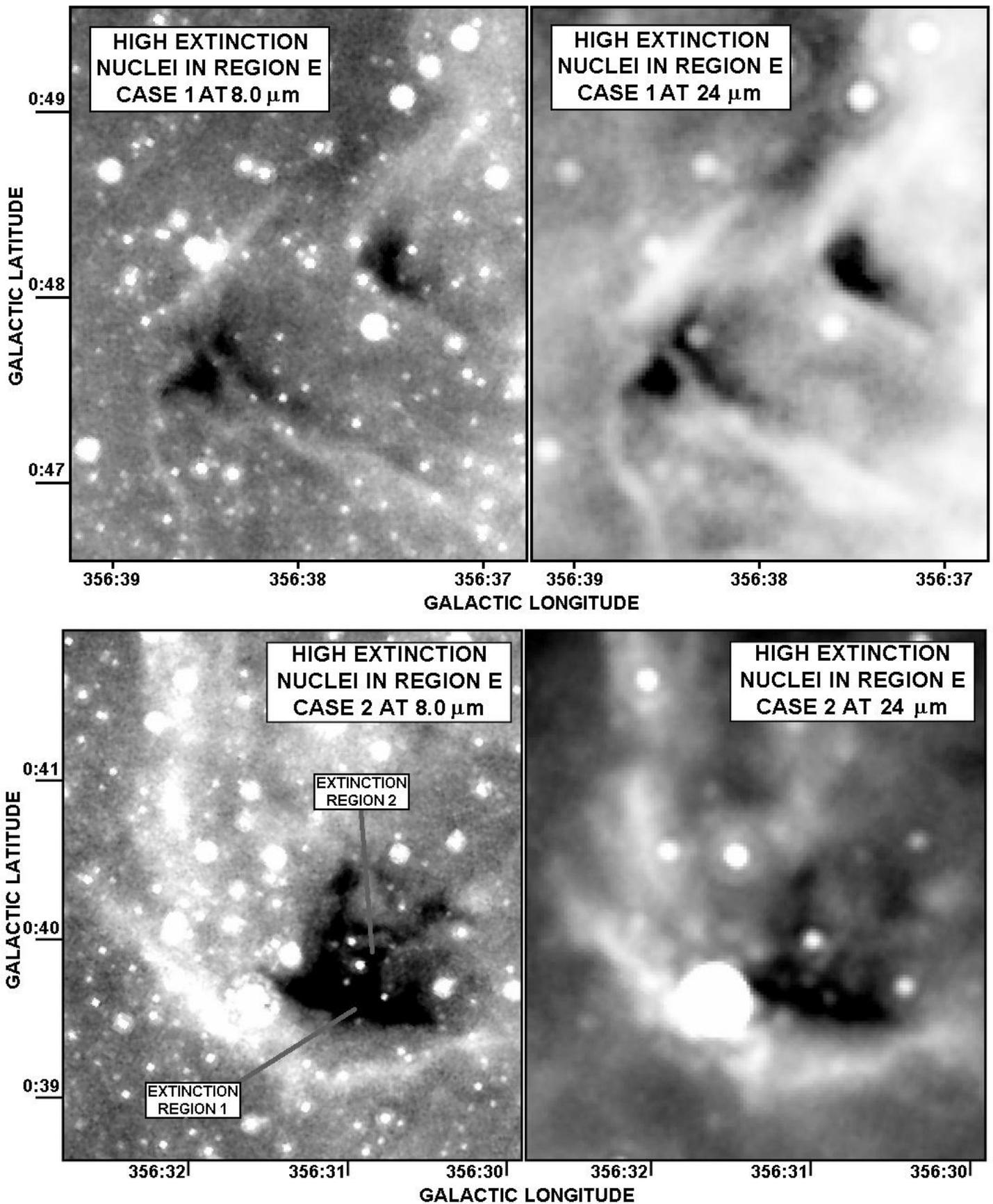

**Figure 14:** Two regions of high extinction cores in region E, illustrated in the IRAC 8.0 μm and MIPSGAL 24 μm imaging. Notice how these regions are extremely compact, and located immediately behind bright frontal structures. The extinction regions 1 and 2 are described in Sect. 5.

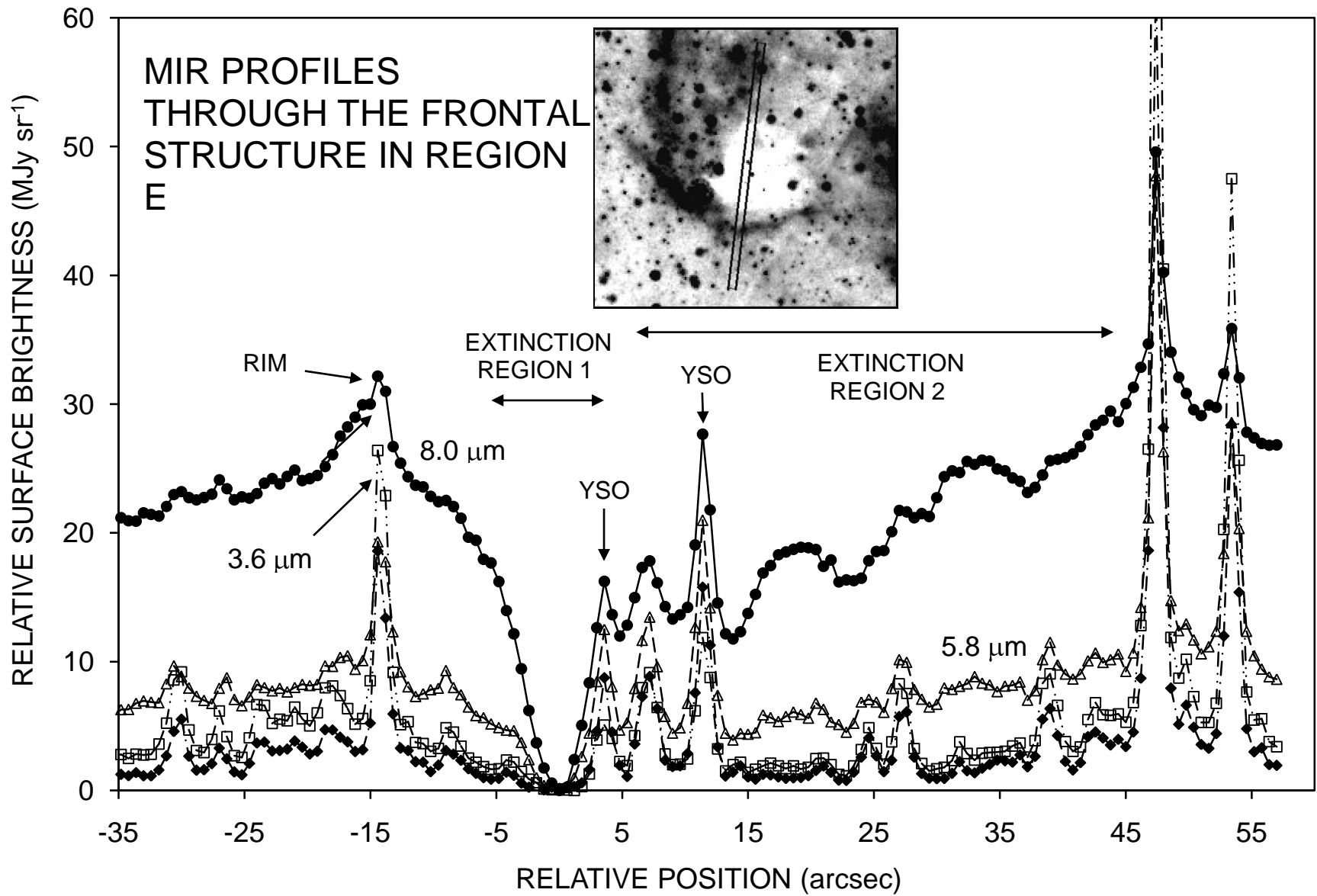

**Figure 15:** As for Fig. 6, but for profiles across the dark extinction core in region E. The core is located close to RP = 0 arcsec, whilst two Class I YSOs are also included within the profiles. The surface brightness parameters ($\Delta 3.6$, $\Delta 4.5$, $\Delta 5.8$, $\Delta 8.0$) are given by (2.1, 2.0, 10.7, 36.88) MJy sr$^{-1}$. We have again identified the extinction regions 1 & 2 described in Sect. 5.

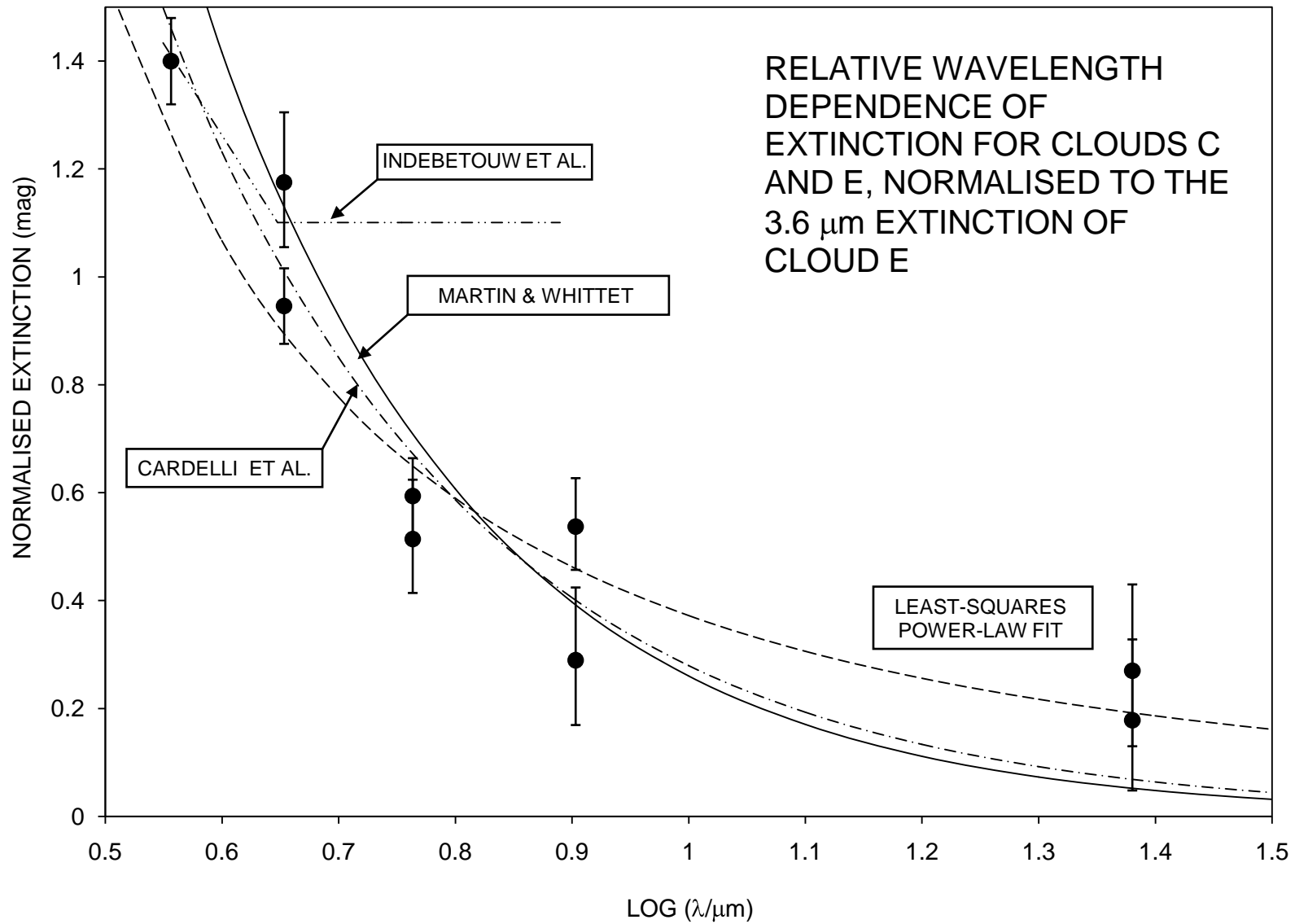

**Figure 16:** The variation of $A_\lambda$ in the high extinction cores of regions C and E, where we also include the extinction trends of Indebetouw et al. (2005), Cardelli et al. (1989) and Martin & Whittet (1990), and a least-squares power-law fit. Note that we have matched the extinctions of the clouds at 3.6 μm by increasing the extinctions for region C by 10%. This makes it easier to see the variations in extinction as a function of wavelength. The actual values of $A_{3.6}$, without such corrections, are $\cong 1.40$ mag for cloud E and $\cong 1.26$ mag for cloud C.

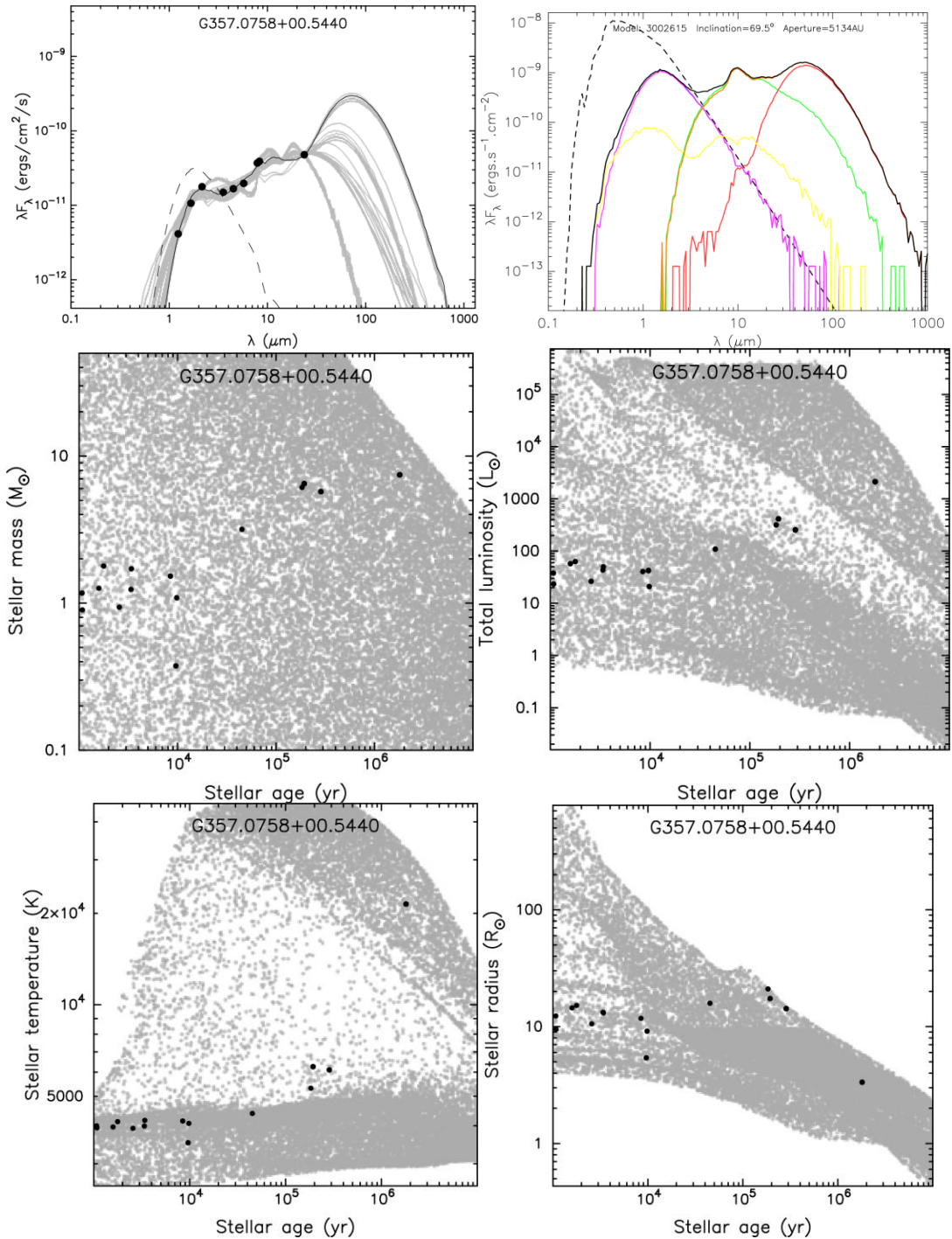

**Figure 17.** Modelling of the Class I YSO located at $l$ = 357.0758°, $b$ = +0.5440°. The top left-hand panel shows photometry taken from 2MASS (J, H, $K_S$), the *SST* (3.6 $\mu$m, 4.5 $\mu$m, 5.8 $\mu$m, 8.0 $\mu$m) and MSX (8.28 $\mu$m), together with YSO fits based upon the modelling of Robitaille et al. (2006). The black curve corresponds to the best-fit model ($\chi_{BEST}^2$ = 10.5), whilst the grey curves represent lower $\chi^2$ models. The dashed curve represents the stellar photospheric continuum, corrected for local dust extinction, but including IS extinction. The more detailed characteristics of the best-fit model are provided in the upper right-hand panel, where the black curve represents the total YSO flux, the stellar photospheric flux is indicated by the dashed line (this is the flux prior to reddening by circumstellar dust); the disk flux is green; the scattered flux is yellow; the envelope flux is red; and the thermal flux is orange. Unless otherwise stated, the results include the effects of circumstellar extinction, but not of IS extinction. They also assume a representative distance of 1 kpc, and an aperture of close to 5 arcsec. Finally, the lower panels show solutions corresponding to the best-fit models, where all of the parameters are represented against the age of the star. The dark bullets represent individual best-fit solutions, whilst grey symbols denote the ranges of parameter explored.

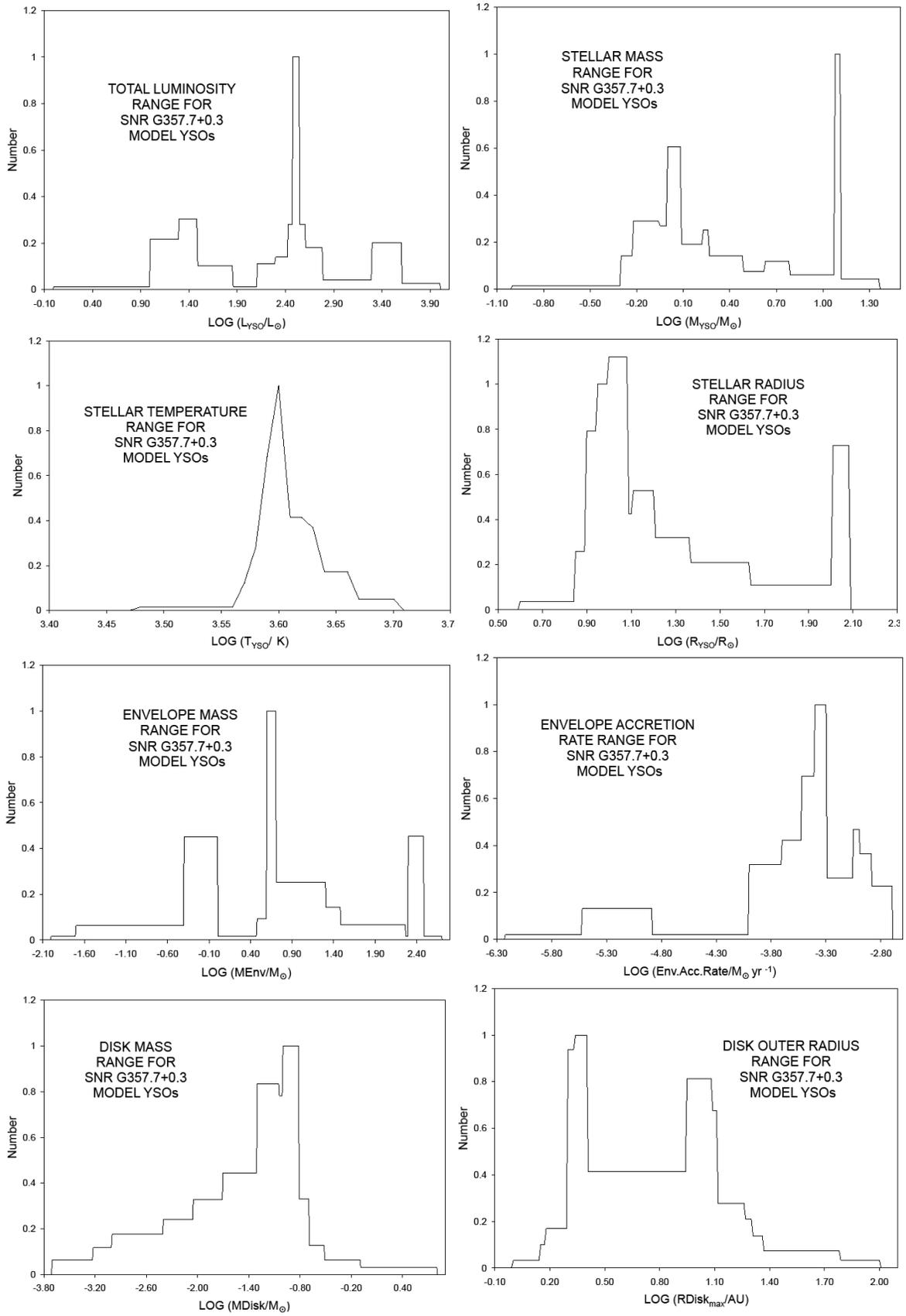

**Figure 18:** Model results for 9 YSOs located close to the frontal structures in regions A→D (see Fig. 2), where we indicate ranges of solution for eight stellar parameters.